\documentclass{article}

\usepackage{graphics}
\usepackage{latexsym}

\newtheorem{theorem}{Theorem}[section]
\newtheorem{conjecture}[theorem]{Conjecture}

\newtheorem{lemma}[theorem]{Lemma}
\newtheorem{claim}[theorem]{Claim}

\newtheorem{algo}{Algorithm}

\def\algorithm#1{\begin{quote}\begin{algo}\
{\sc (#1)} \rm }
\def\endalgorithm{\end{algo}\end{quote}}

\def\vertex{\circle*{0.4}}

\begin{document}

\begin{center}

{\Large {\bf Algorithms for perfectly contractile graphs}}

\vspace{2em}

{\large{\bf   Fr\'{e}d\'{e}ric   Maffray\footnote{C.N.R.S.},   Nicolas
Trotignon\footnote{Supported  by  Universit\'e  Pierre Mendes  France,
Grenoble.}}}

{\small  Laboratoire  Leibniz-IMAG,  46  Avenue  F\'{e}lix  Viallet,\\
38031~Grenoble~Cedex,       France.\\       (frederic.maffray@imag.fr,
nicolas.trotignon@imag.fr)}

February 10, 2005

\end{center}
\begin{quote}
\small {\bf Abstract.} We consider the class ${\cal A}$ of graphs that
contain no odd hole, no antihole of length at least $5$, and no
``prism'' (a graph consisting of two disjoint triangles with three
disjoint paths between them) and the class ${\cal A}'$ of graphs that
contain no odd hole, no antihole of length at least $5$, and no odd
prism (prism whose three paths are odd).  These two classes were
introduced by Everett and Reed and are relevant to the study of
perfect graphs.  We give polynomial-time recognition algorithms for
these two classes.  We proved previously that every graph $G\in{\cal
A}$ is ``perfectly contractile'', as conjectured by Everett and Reed
[see the chapter ``Even pairs'' in the book {\it Perfect Graphs},
J.L.~Ram\'{\i}rez-Alfons\'{\i}n and B.A.~Reed, eds., Wiley
Interscience, 2001].  The analogous conjecture concerning graphs in
${\cal A}'$ is still open.
\end{quote}

\section{Introduction}

A graph $G$ is \emph{perfect} if every induced subgraph $G'$ of $G$
satisfies $\chi(G')=\omega(G')$, where $\chi(G')$ is the chromatic
number of $G'$ and $\omega(G')$ is the maximum clique size in $G'$.
Berge {\cite{ber60,ber61,ber85}} introduced perfect graphs and
conjectured that \emph{a graph is perfect if and only if it does not
contain as an induced subgraph an odd hole or an odd antihole} (the
Strong Perfect Graph Conjecture), where a \emph{hole} is a chordless
cycle with at least four vertices and an \emph{antihole} is the
complement of a hole.  We follow the tradition of calling \emph{Berge
graph} any graph that contains no odd hole and no odd antihole.  The
Strong Perfect Graph Conjecture was the objet of much research (see
the book \cite{ramree01}), until it was finally proved by Chudnovsky,
Robertson, Seymour and Thomas \cite{CRST2002}: \emph{Every Berge graph
is perfect}.  Moreover, Chudnovsky, Cornu\'ejols, Liu, Seymour and
Vu\v{s}kovi\'c \cite{CCLSV2002,CLV2002,CS2002} gave polynomial-time
algorithms to decide if a graph is Berge.

Despite those breakthroughs, some conjectures about Berge graphs
remain open.  An \emph{even pair} in a graph $G$ is a pair of
non-adjacent vertices such that every chordless path between them has
even length (number of edges).  Given two vertices $x,y$ in a graph
$G$, the operation of \emph{contracting} them means removing $x$ and
$y$ and adding one vertex with edges to every vertex of $G\setminus
\{x,y\}$ that is adjacent in $G$ to at least one of $x,y$; we denote
by $G/xy$ the graph that results from this operation.  Fonlupt and
Uhry \cite{fonuhr82} proved that \emph{if $G$ is a perfect graph and
$\{x,y\}$ is an even pair in $G$, then the graph $G/xy$ is perfect and
has the same chromatic number as $G$}.  In particular, given a
$\chi(G/xy)$-coloring $c$ of the vertices of $G/xy$, one can easily
obtain a $\chi(G)$-coloring of the vertices of $G$ as follows: keep
the color for every vertex different from $x,y$; assign to $x$ and $y$
the color assigned by $c$ to the contracted vertex.  This idea could
be the basis for a conceptually simple coloring algorithm for Berge
graphs: as long as the graph has an even pair, contract any such pair;
when there is no even pair find a coloring $c$ of the contracted graph
and, applying the procedure above repeatedly, derive from $c$ a
coloring of the original graph.  The polynomial-time algorithm for
recognizing Berge graphs mentioned at the end of the preceding
paragraph can be used to detect an even pair in a Berge graph $G$;
indeed, two non-adjacent vertices $a,b$ form an even pair in $G$ if
and only if the graph obtained by adding a vertex adjacent only to $a$
and $b$ is Berge.  The problem of deciding if a graph contains an even
pair is NP-hard in general graphs \cite{bien}.  Given a Berge graph
$G$, one can try to color its vertices by keeping contracting even
pairs until none can be found.  Then some questions arise: what are
the Berge graphs with no even pair?  What are, on the contrary, the
graphs for which a sequence of even-pair contractions leads to graphs
that are easy to color?

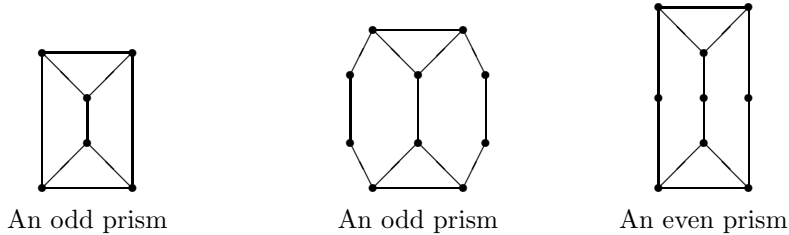
\begin{figure}[htb]
\unitlength=0.3cm
\begin{center}\begin{tabular}{ccc}
\begin{picture}(6,9)
\multiput(0,0)(0,6){2}{\vertex}
\multiput(2,2)(0,2){2}{\vertex}
\multiput(4,0)(0,6){2}{\vertex}
\multiput(0,0)(0,6){2}{\line(1,0){4}}
\multiput(0,0)(4,0){2}{\line(0,1){6}}
\put(2,2){\line(0,1){2}}
\multiput(0,0)(2,4){2}{\line(1,1){2}}
\multiput(0,6)(2,-4){2}{\line(1,-1){2}}
\put(2,-1.5){\makebox(0,0){An odd prism}}
\end{picture}
\quad\quad\quad & \quad\quad\quad
\begin{picture}(6,9)
\multiput(1,0)(4,0){2}{\vertex}
\multiput(0,2)(3,0){3}{\vertex}
\multiput(0,5)(3,0){3}{\vertex}
\multiput(1,7)(4,0){2}{\vertex}
\multiput(1,0)(0,7){2}{\line(1,0){4}}
\multiput(0,2)(3,0){3}{\line(0,1){3}}
\multiput(1,0)(2,5){2}{\line(1,1){2}}
\multiput(1,7)(2,-5){2}{\line(1,-1){2}}
\multiput(0,5)(5,-5){2}{\line(1,2){1}}
\multiput(0,2)(5,5){2}{\line(1,-2){1}}
\put(3,-1.5){\makebox(0,0){An odd prism}}
\end{picture}
\quad\quad\quad & \quad\quad\quad
\begin{picture}(6,9)
\multiput(0,0)(0,4){3}{\vertex}
\multiput(2,2)(0,2){3}{\vertex}
\multiput(4,0)(0,4){3}{\vertex}
\multiput(0,0)(0,8){2}{\line(1,0){4}}
\multiput(0,0)(4,0){2}{\line(0,1){8}}
\put(2,2){\line(0,1){4}}
\multiput(0,0)(2,6){2}{\line(1,1){2}}
\multiput(0,8)(2,-6){2}{\line(1,-1){2}}
\put(2,-1.5){\makebox(0,0){An even prism}}
\end{picture}
\end{tabular}\end{center}
\vspace{0.6cm}
\caption{Some prisms}\label{fig:prisms}
\end{figure}

As a first step towards getting a better grasp on these questions,
Bertschi \cite{ber90} proposed the following definitions.  A graph $G$
is \emph{even-contractile} if either $G$ is a clique or there exists a
sequence $G_0, \ldots, G_k$ of graphs such that $G=G_0$, for $i=0,
\ldots, k-1$ the graph $G_i$ has an even pair $\{x_i, y_i\}$ such that
$G_{i+1}=G_i/x_iy_i$, and $G_k$ is a clique.  A graph $G$ is
\emph{perfectly contractile} if every induced subgraph of $G$ is
even-contractile.  Perfectly contractile graphs include many classical
families of perfect graphs, such as Meyniel graphs, weakly chordal
graphs, perfectly orderable graphs, see \cite{epsbook}.  Everett and
Reed proposed a conjecture aiming at a characterization of perfectly
contractile graphs.  To understand it, one more definition is needed:
say that a graph is a \emph{prism} if it consists of two
vertex-disjoint triangles (cliques of size $3$) $\{a_1, a_2, a_3\}$,
$\{b_1, b_2, b_3\}$, with three vertex-disjoint paths $P_1, P_2, P_3$
between them, such that for $i=1, 2, 3$ path $P_i$ is from $a_i$ to
$b_i$, and with no other edge than those in the two triangles and in
the three paths.  We may also say that the three paths $P_1, P_2, P_3$
\emph{form} the prism.  Say that a prism is odd (or even) if all three
paths have odd length (respectively all have even length).  See
Figure~\ref{fig:prisms}.

Define two classes $\cal A$, ${\cal A}'$ of graphs as follows:
\begin{itemize}
\item
${\cal  A}$ is  the class  of graphs  that do  not contain  odd holes,
antiholes of length at least $5$, or prisms.
\item
${\cal A}'$  is the  class of  graphs that do  not contain  odd holes,
antiholes of length at least $5$, or odd prisms.
\end{itemize}
Clearly ${\cal A}\subset {\cal A}'$.

\begin{conjecture}[Everett and Reed \cite{epsbook,ree93}]
\label{conj:pc}
A graph is perfectly contractile if  and only if it is in class ${\cal
A}'$.
\end{conjecture}
The if part of this conjecture remains open.  The only if part is not
hard to establish, but it requires some careful checking; this was
done formally in \cite{linmafree97}.  A weaker form of this conjecture
was also proposed by Everett and Reed; that statement is now a
theorem: %
\begin{theorem}[Maffray and Trotignon \cite{maftro02}]
\label{thm:maftro}
If $G$ is  a graph in class ${\cal  A}$ and $G$ is not  a clique, then
$G$ has  an even pair whose  contraction yields a graph  in ${\cal A}$
(and so $G$ is perfectly contractile).
\end{theorem}

The preceding conjecture and theorem suggest that it may be
interesting to recognize the classes $\cal A$ and ${\cal A}'$ in
polynomial time; this is the aim of this manuscript.

In order to decide if a graph is in class $\cal A$, it would suffice
to decide separately if it is Berge, if it has an antihole of length
at least $5$, and if it contains a prism.  The first question,
deciding if a graph is Berge, is now settled \cite{CCLSV2002, CS2002,
CLV2002}.  In Section~\ref{sec:berge} we will find it convenient for
our purpose to give a summary of the polynomial time algorithm from
\cite{CCLSV2002, CS2002} that solves this problem.  The second
question is not hard: to decide if a graph $G$ contains a hole of
length at least $5$, it suffices to test, for every chordless path
$a$-$b$-$c$, whether $a$ and $c$ are in the same connected component
of the subgraph of $G$ obtained by removing the vertices of $N(a)\cap
N(c)$ and those of $N(b)\setminus\{a, c\}$.  This takes time
$O(|V(G)|^5)$.  To decide if a graph contains an antihole of length at
least $5$, we need only apply this algorithm on its complementary
graph.  However, the third question, to decide if a graph contains a
prism, turns out to be NP-complete; this is established in
Section~\ref{sec:npc} below.  Likewise, we will see that it is
NP-complete to decide if a graph contains an odd prism.  Thus we
cannot solve the recognition problem for class $\cal A$ (or for class
${\cal A}'$) in the fashion that is suggested at the beginning of this
paragraph.  Instead, we will adapt the Berge graph recognition
algorithm to our purpose.  This is done in
Sections~\ref{sec:pyrpri}--\ref{sec:aprime}.


\section{Recognizing Berge graphs}
\label{sec:berge}

We give here a brief outline of the Berge graph recognition algorithm
which follows from \cite{CCLSV2002} and \cite{CS2002}.  Given a graph
$G$ and a hole $C$ in $G$, say that a vertex $x\in V(G)\setminus V(C)$
is a \emph{major neighbour} of $C$ if the set $N(x)\cap V(C)$ is not
included in a $3$-vertex subpath of $C$.  Say that set $X\subseteq
V(G)$ is a \emph{cleaner} for the hole $C$ if $X$ contains all the
major neighbours of $C$ and $X\cap V(C)$ is included in a $3$-vertex
subpath of $C$.  The algorithm is based on the results summarized in
the following theorem.

\begin{theorem}[\cite{CCLSV2002,CS2002}]\ \\
1.  There  exist  five types  of  configurations  (graphs), types  T1,
\ldots, T5, such  that, for $i=1, \ldots, 5$, we have:  (a) if a graph
$G$ contains  a configuration  of type  T$i$ then $G$  is not  a Berge
graph,  and (b)  there exists  a polynomial  time algorithm  A$i$ that
decides if a graph contains a configuration of type T$i$.

2.  There is a polynomial-time algorithm which, given a graph $G$ that
does  not contain  a configuration  of type  T$i$ ($i=1,  \ldots, 5$),
returns a  family $\cal F$ of  $|V(G)|^5$ subsets of  $V(G)$ such that
for any  shortest odd hole $C$  of $G$, some  member of $\cal F$  is a
cleaner for $C$.

3.  There is a polynomial-time algorithm which, given a graph $G$ that
does not contain a configuration of type T$i$ ($i=1, \ldots, 5$) and
the family $\cal F$ produced by step 2, decides if $G$ contains an odd
hole (and if it does, returns a shortest odd hole of $G$).
\end{theorem}

We will not give the definition of all five types of configurations,
but we recall from \cite{CCLSV2002,CS2002} that, for $i=1, \ldots, 5$,
the complexity of algorithm A$i$ is respectively $O(|V(G)|^5)$,
$O(|V(G)|^6)$, $O(|V(G)|^6)$, $O(|V(G)|^6)$, $O(|V(G)|^9)$.  We need
to dwell on the configuration of type T$5$, which is called a
\emph{pyramid} in~\cite{CS2002}.  A pyramid is a graph that consists
in three pairwise adjacent vertices $b_1, b_2, b_3$ (called the
triangle vertices of the pyramid), a fourth vertex $a$ (called the
apex of the pyramid), and three chordless paths $P_1, P_2, P_3$ such
that:
\begin{itemize}
\item
For $i=1, 2, 3$, path $P_i$ is between $a$ and $b_i$;
\item
For  $1\le i<j\le 3$,  $V(P_i)\cap V(P_j)=\{a\}$  and $b_ib_j$  is the
only edge between $V(P_i)\setminus \{a\}$ and $V(P_j)\setminus \{a\}$;
\item
$a$ is adjacent to at most one of $b_1, b_2, b_3$.
\end{itemize}
We may say that the three paths $P_1, P_2, P_3$ \emph{form} a pyramid.
It is easy to see that a pyramid contains an odd hole (since two of
the paths $P_1, P_2, P_3$ have the same parity, the union of their
vertex sets induce an odd hole); so Berge graphs do not contain
pyramids.

The pyramid-testing algorithm from~\cite{CS2002} is the slowest
algorithm in Step 1 of the Berge graph recognition algorithm.  The
algorithm of Step 2 has complexity $O(|V(G)|^6)$ \cite{CCLSV2002}, and
the algorithm of Step 3 has complexity $O(|V(G)|^9)$ \cite{CS2002}.
Testing if a graph $G$ is Berge can be done by running the algorithms
described in the previous theorem on $G$ and on its complementary
graph $\overline{G}$.  Thus the total complexity is $O(|V(G)|^9)$.


\section{Recognizing pyramids and prisms}
\label{sec:pyrpri}

We present a polynomial-time algorithm that decides if a graph
contains a pyramid or a prism.  This algorithm has the same flavor as
the pyramid-testing algorithm from \cite{CS2002}.  We describe this
algorithm now.

If a graph contains  a pyramid or a prism, it contains  a pyramid or a
prism that is \emph{smallest} in the sense that there is no pyramid or
prism induced by strictly fewer vertices.  Smallest pyramids or prisms
have properties that make them easier to handle.  These properties
are expressed in the next two lemmas.

\begin{lemma}
\label{lem:kpyr}
Let $G$ be  a graph.  Let $K$  be a smallest pyramid or  prism in $G$.
Suppose that $K$  is a pyramid, formed by paths  $P_1, P_2, P_3$, with
triangle $\{b_1,  b_2, b_3\}$ and apex  $a$.  Let $R_1$  be a shortest
path from  $b_1$ to  $a$ whose interior  vertices are not  adjacent to
$b_2$ or  $b_3$.  Then the subgraph induced  by $V(R_1)\cup V(P_2)\cup
V(P_3)$ is a smallest pyramid or prism in $G$.
\end{lemma}
\emph{Proof.} Note that $|V(R_1)|\le |V(P_1)|$ since $P_1$ is a path
from $b_1$ to $a$ whose interior vertices are not adjacent to $b_2$ or
$b_3$.  Let $P$ be the path induced by $(V(P_2) \setminus \{b_2\})\cup
(V(P_3) \setminus \{b_3\})$.  If no vertex of $R_1\setminus \{a\}$ has
any neighbour in $P\setminus\{a\}$, then $R_1, P_2, P_3$ form a
pyramid in $G$, and its number of vertices is note larger than
$|V(K)|$, so the lemma holds.  So we may assume that some vertex $c$
of $R_1\setminus \{a\}$ has a neighbour in $P\setminus\{a\}$, and we
choose $c$ closest to $b_1$ along $R_1$.  Recall that $c$ is not
adjacent to $b_2$ or $b_3$, by the definition of $R_1$.  For $j=2, 3$,
let $b'_j$ be the neighbour of $b_j$ along $P_j$ (so $b'_2, b'_3$ are
the ends of $P$) and let $c_j$ be the neighbour of $c$ closest to
$b'_j$ along $P$.

Suppose $c_2 = c_3$.  We have $c_3 \neq a$ since $c$ has a neighbour
along $P\setminus \{a\}$.  Then the three chordless paths
$c_2$-$c$-$R_1$-$b_1$, $c_2$-$P$-$b_2$, $c_2$-$P$-$b_3$ form a pyramid
with triangle $\{b_1, b_2, b_3\}$ and apex $c_2$; this pyramid is
strictly smaller than $K$, because it is included in $(V(R_1)\setminus
\{a\})\cup V(P_2)\cup V(P_3)$, a contradiction.  So $c_2\neq c_3$.  If
$c_2, c_3$ are not adjacent, then the three chordless paths
$c$-$R_1$-$b_1$, $c$-$c_2$-$P$-$b_2$, $c$-$c_3$-$P$-$b_3$ form a
pyramid with triangle $\{b_1, b_2, b_3\}$ and apex $c$; again this
pyramid has strictly fewer vertices than $K$, a contradiction.  So
$c_2, c_3$ are adjacent.  Then the three chordless paths
$c$-$R_1$-$b_1$, $c_2$-$P$-$b_2$ and $c_3$-$P$-$b_3$ form a prism
$K'$, with triangles $\{b_1, b_2, b_3\}$ and $\{c, c_2, c_3\}$.  If
$a\notin \{c_2, c_3\}$ then $K'$ is smaller than $K$, a contradiction.
So $a\in \{c_2, c_3\}$ and the prism $K'$ has the same size as $K$, so
the lemma holds.  $\Box$

\begin{lemma}
\label{lem:kpri}
Let $G$ be  a graph.  Let $K$  be a smallest pyramid or  prism in $G$.
Suppose that  $K$ is a  prism, formed by  paths $P_1, P_2,  P_3$, with
triangles $\{a_1,  a_2, a_3\}$ and  $\{b_1, b_2, b_3\}$, so  that, for
$i=1, 2, 3$, path $P_i$ is from $a_i$ to $b_i$.  Then:
\begin{itemize}
\item
If $R_1$ is any shortest path from $a_1$ to $b_1$ whose interior
vertices are not adjacent to $b_2$ or $b_3$, then $R_1, P_2, P_3$ form
a prism of size $|V(K)|$ in $G$, with triangles $\{a_1, a_2, a_3\}$
and $\{b_1, b_2, b_3\}$.
\item
If $R_2$ is any shortest path from $a_1$ to $b_2$ whose interior
vertices are not adjacent to $b_1$ or $b_3$, then either the three
paths $P_1, R_2\setminus a_1, P_3$ form a smallest prism in $G$, or
the three paths $P_1, R_2, P_3+a_1$ form a pyramid of size $|V(K)|$ in
$G$, with triangle $\{b_1, b_2, b_3\}$ and apex $a_1$.
\end{itemize}
\end{lemma}

\emph{Proof.} Let  us prove  the first item  of the lemma.   Note that
$|V(R_1)|\le |V(P_1)|$ since $P_1$ is a path from $a_1$ to $b_1$ whose
interior vertices are not adjacent to  $b_2$ or $b_3$.  Let $P$ be the
path  induced by  $(V(P_2) \setminus  \{b_2\}) \cup  (V(P_3) \setminus
\{b_3\})$.  If no  interior vertex of $R_1$ is  adjacent to any vertex
of $V(P)$,  then the three paths $R_1,  P_2, P_3$ form a  prism in $G$
whose  size is  not larger  than the  size  of $K$,  so it  must be  a
smallest prism and the lemma holds.  So we may assume that there is an
interior vertex  $c$ of $R_1$  that has a  neighbour in $V(P)$  and we
choose $c$ closest to $b_1$ along  $R_1$.  For $j=2, 3$, let $b'_j$ be
the neighbour  of $b_j$ along $P_j$  (so $b'_2, b'_3$ are  the ends of
$P$) and  let $c_j$ be  the neighbour of  $c$ closest to  $b'_j$ along
$P$.

Suppose $c_2 = c_3$.  Then the three paths $c_2$-$c$-$R_1$-$b_1$,
$c_2$-$P$-$b_2$, $c_2$-$P$-$b_3$ form a pyramid with triangle $\{b_1,
b_2, b_3\}$ and apex $c_2$; this pyramid is strictly smaller than $K$
(since $|V(R_1\setminus\{a\})|< |V(P_1)|$), a contradiction.  So
$c_2\neq c_3$.  If $c_2, c_3$ are adjacent, then the three paths
$c$-$R_1$-$b_1$, $c_2$-$P$-$b_2$, $c_3$-$P$-$b_3$ form a prism, with
triangles $\{b_1, b_2, b_3\}$ and $\{c, c_2, c_3\}$, that is strictly
smaller than $K$, a contradiction.  So $c_2, c_3$ are not adjacent.
But then the three paths $c$-$R_1$-$b_1$, $c$-$c_2$-$P$-$b_2$,
$c$-$c_3$-$P$-$b_3$ form a pyramid with triangle $\{b_1, b_2, b_3\}$,
apex $c$, and this pyramid is strictly smaller than $K$, a
contradiction.  So the first item is proved.

Now we  prove the  second item of  the lemma.  Note  that $|V(R_2)|\le
|V(P_2)|+1$  since $P_2+a_1$  is  a  path from  $a_1$  to $b_2$  whose
interior vertices are not adjacent to  $b_2$ or $b_3$.  Let $P$ be the
path   induced  by  $(V(P_1)\setminus\{b_1\})\cup   (V(P_3)  \setminus
\{b_3\})$.   If no  interior  vertex  of $R_2$  has  any neighbour  in
$V(P\setminus a_1)$ then $P_1, R_2,  P_3+a_1$ form a pyramid, which is
not  larger than  $K$; so  it is  a smallest  pyramid and  the theorem
holds.  Now assume that some  interior vertex of $R_2$ has a neighbour
in $V(P)$,  and choose the  vertex $c$ that  has this property  and is
closest to $b_2$.  For $i= 1, 3$, let $b'_i$ be the neighbour of $b_i$
along $P_i$ (so $b'_1, b'_3$ are the ends of $P$) and let $c_i$ be the
neighbour of $c$ along $P$ that is closest to $b'_i$.

Suppose $c_1 = c_3$.  Then $c_1 \neq a_1$ since $c$ has a neighbour in
$V(P\setminus a_1)$.  Then the three paths $c_1$-$c$-$R_2$-$b_2$,
$c_1$-$P$-$b_1$, $c_1$-$P$-$b_3$ from a pyramid with triangle $\{b_1,
b_2, b_3\}$ and apex $c_1$.  This pyramid is strictly smaller than
$K$, a contradiction.  So $c_1\neq c_3$.  If $c_1, c_3$ are not
adjacent, then the three paths $c$-$R_2$-$b_2$, $c$-$c_1$-$P$-$b_1$,
$c$-$c_3$-$P$-$b_3$ form a pyramid with triangle $\{b_1, b_2, b_3\}$
and apex $c$; this pyramid has size strictly smaller than $K$, a
contradiction.  So $c_1, c_3$ are adjacent.  Then the three paths
$c$-$R_2$-$b_2$, $c_1$-$P$-$b_1$, $c_3$-$P$-$b_3$ form a prism $K'$,
with triangles $\{b_1, b_2, b_3\}$ and $\{c, c_1, c_3\}$.  If
$a_1\notin \{c_1, c_3\}$ then this prism is strictly smaller than $K$,
a contradiction.  So $a_1\in \{c_1, c_3\}$ and $K'$ has the same size
as $K$, and the lemma holds.  This completes the proof of the lemma.
$\Box$

On the basis  of the preceding lemmas we can  present an algorithm for
testing if a graph contains a pyramid or a prism.

\begin{algorithm}{Detection of a pyramid or prism}
\label{alg:pyrpri}

\emph{Input:} A graph $G$.

\emph{Output:} An  induced pyramid  or prism of  $G$, if  $G$ contains
any;  else  the negative  answer  ``$G$  contains  no pyramid  and  no
prism.''

\emph{Method:} For every  quadruple $a, b_1, b_2, b_3$  of vertices of
$G$  such  that $b_1,  b_2,  b_3$ are  pairwise  adjacent  and $a$  is
adjacent to  at most one  of them, do:  Compute a shortest  path $P_1$
from $a$  to $b_1$ whose interior  vertices are not  adjacent to $b_2,
b_3$, if any.  Compute paths  $P_2$ and $P_3$ similarly.  If the three
paths  $P_1, P_2, P_3$  exist, and  if $V(P_1)\cup  V(P_2)\cup V(P_3)$
induces a  pyramid or a prism,  then return this subgraph  of $G$, and
stop.

If no quadruple has produced a pyramid or a prism, return the negative
answer.

\emph{Complexity:} $O(|V(G)|^6)$.
\end{algorithm}

\emph{Proof of correctness.}  If $G$ contains no pyramid  and no prism
then  clearly   the  algorithm   will  return  the   negative  answer.
Conversely, suppose that  $G$ contains a pyramid or  a prism.  Let $K$
be a smallest  pyramid or prism.  Let $b_1, b_2,  b_3$ be the vertices
of a  triangle of $K$, and  let $a$ be such  that if $K$  is a pyramid
then $a$ is its apex and if $K$ is a prism then $a$ is a vertex of the
other triangle of $K$.  When our algorithm considers the quadruple $a,
b_1, b_2, b_3$, it will find paths $P_1, P_2, P_3$ since some paths in
$K$  do have  the required  properties.  Then,  three  applications of
lemmas~\ref{lem:kpyr} and~\ref{lem:kpri} imply that $P_1, P_2, P_3$ do
form a pyramid  or a prism of $G$.  So the  algorithm will detect this
subgraph.

Complexity analysis:  Testing all quadruples  take time $O(|V(G)|^4)$.
For each  quadruple, finding the three paths  takes time $O(|V(G)|^2)$
and checking  that the  corresponding subgraph is  a pyramid  or prism
takes time $O(|V(G)|)$.  Thus the overall complexity is $O(|V(G)|^6)$.
$\Box$

We  now  show  how the  results  of  the  preceding algorithm  can  be
performed a little bit faster with a simple trick.

\begin{lemma}
\label{lem:3sorties}
Let $H$  be a  connected graph  and let $V_1,  V_2, V_3$  be non-empty
subsets of  $V(H)$.  Then  $H$ has an  induced subgraph $F$  such that
either:
\begin{enumerate}
\item\label{fp} $F$ is a chordless path such that, up to a permutation
of $V_1, V_2, V_3$, one end of $F$ is in $V_1$, the other is in $V_3$,
some vertex  of $F$ is in  $V_2$ and no  interior vertex of $F$  is in
$V_1\cup V_3$;
\item\label{fc} $F$ consists of  three chordless paths $F_1, F_2, F_3$
of length at least  $1$ such that: for $i=1, 2, 3$,  $F_i$ is from $f$
to  $v_i$   and  $v_i\in  V_i$;  for  $1\le   i<j\le  3$,  $V(F_i)\cap
V(F_j)=\{f\}$  and  there is  no  edge  between  $F_i\setminus f$  and
$F_j\setminus f$; and $F\setminus\{v_1, v_2, v_3\}$ contains no vertex
of $V_1\cup V_2\cup V_3$;
\item\label{ft} $F$ consists  of three vertex-disjoint chordless paths
$F_1, F_2, F_3$  (possibly of length $0$) such that:  for $i=1, 2, 3$,
$F_i$ is  from $w_i$  to $v_i$ and  $v_i\in V_i$; vertices  $w_1, w_2,
w_3$  are pairwise  adjacent; for  $1\le i<j\le  3$ there  is  no edge
between  $F_i$ and  $F_j$ other  than $w_iw_j$;  and $F\setminus\{v_1,
v_2, v_3\}$ contains no vertex of $V_1\cup V_2\cup V_3$.
\end{enumerate}
\end{lemma}
\emph{Proof.} Let $P$ be a shortest  path in $H$ such that $P$ has one
end in $V_1$  and the other in $V_3$; let $v_1\in  V_1, v_3\in V_3$ be
the ends of $P$.  Thus no  interior vertex of $P$ is in $V_1\cup V_3$.
If $P$ contains a vertex of $V_2$ then we have outcome~\ref{fp} of the
lemma with $F=P$.  Therefore let us assume that $P$ contains no vertex
of $V_2$.  Let $Q$  be a shortest path such that one  end $v_2$ of $Q$
is in $V_2$ and the other end  $v$ of $Q$ has a neighbour on $P$.  Let
$w, x$ be the  neighbours of $v$ on $P$ that are  closest to $v_1$ and
$v_3$ respectively.  Note that  $Q\setminus v_2$ contains no vertex of
$V_2$  by the definition  of $Q$.   If $Q$  contains vertices  of both
$V_1, V_3$ then  some subpath $F$ of $Q$ contains  vertices of each of
$V_1,  V_2,  V_3$  and is  minimal  with  this  property, and  so  $F$
satisfies outcome~\ref{fp} of the  lemma.  If $Q$ contains vertices of
$V_1$ and not of $V_3$, then $v_3$-$P$-$x$-$v$-$Q$-$v_2$ is a path $F$
that  satisfies outcome~\ref{fp}.   A similar  outcome happens  if $Q$
contains vertices  of $V_3$ and not  of $V_1$.  So we  may assume that
$Q$ contains no vertex of $V_1\cup V_3$.

Suppose $w=x$.  If $x\in\{v_1, v_3\}$, we have outcome~\ref{fp} with
$F= P+Q$.  If $x\notin\{v_1, v_3\}$, the three paths $x$-$P$-$v_1$,
$x$-$v$-$Q$-$v_2$, $x$-$P$-$v_3$ form a subgraph $F$ that satisfies
outcome~\ref{fc}.  Now suppose that $w,x$ are different and not
adjacent.  If $v=v_2$, then $v_1$-$P$-$w$-$v_2$-$x$-$P$-$v_3$ is a
path $F$ that satisfies outcome~\ref{fp}.  If $v\not=v_2$, the three
paths $v$-$w$-$P$-$v_1$, $v$-$Q$-$v_2$, $v$-$x$-$P$-$v_3$ form a
subgraph $F$ that satisfies outcome~\ref{fc}.  Finally, suppose that
$w,x$ are different and adjacent.  Then the three paths $w$-$P$-$v_1$,
$v$-$Q$-$v_2$, $x$-$P$-$v_3$ form a subgraph $F$ that satisfies the
properties of outcome~\ref{ft}.  This completes the proof of the
lemma.  $\Box$

Now we can give an algorithm:

\begin{algorithm}{Detection of a pyramid or prism}
\label{alg:pyrpri2}

\emph{Input:} A graph $G$.

\emph{Output:}  The positive  answer  ``$G$ contains  a  pyramid or  a
prism'' if it does; else the negative answer ``$G$ contains no pyramid
and no prism.''

\emph{Method:} For every triple $b_1, b_2, b_3$ of vertices of $G$
such that $b_1, b_2, b_3$ are pairwise adjacent, do: \\ Step 1.
Compute the set $X_1$ of those vertices of $V(G)$ that are adjacent to
$b_1$ and not adjacent to $b_2$ or $b_3$, and the similar sets $X_2,
X_3$, and compute the set $X$ of those vertices of $V(G)$ that are not
adjacent to any of $b_1, b_2, b_3$.  If some vertex of any $X_i$ has a
neighbour in each of the other two $X_j$'s, return the positive answer
and stop.  Else:\\ Step 2.  Compute the connected components of $X$ in
$G$.\\ Step 3.  For each component $H$ of $X$, and for $i=1, 2, 3$, if
some vertex of $H$ has a neighbour in $X_i$ then mark $H$ with label
$i$.  If any component $H$ of $X$ gets the three labels $1, 2, 3$,
return the positive answer and stop.\\ If no triple yields the
positive answer, return the negative answer.

\emph{Complexity:} $O(|V(G)|^5)$.
\end{algorithm}

\emph{Proof of correctness.} Suppose that $G$ contains a pyramid or a
prism $K$.  Let $b_1, b_2, b_3$ be the vertices of a triangle of $K$,
and for $i=1, 2, 3$ let $c_i$ be the neighbour of $b_i$ in $K
\setminus\{b_1, b_2, b_3\}$.  The algorithm will place the three
vertices $c_1, c_2, c_3$ in the sets $X_1, X_2, X_3$ respectively, one
vertex in each set.  If $K$ has only six vertices, the algorithm will
find that one of the $c_i$'s is adjacent to the other two, so it will
return the positive answer at the end of Step 1.  If $K$ has at least
seven vertices, then the algorithm will place the vertices of
$K'=K\setminus\{b_1, b_2, b_3, c_1, c_2, c_3\}$ in $X$; at Step 2
these vertices will all be in one component of $X$ since $K'$ is
connected, and at Step 3 this component with get the three labels $1,
2, 3$ since $K'$ contains a neighbour of $c_i$ for each $i=1, 2, 3$,
so the algorithm will return the positive answer.

Conversely, suppose that the algorithm returns the positive answer
when it is examining a triple $\{b_1, b_2, b_3\}$ that induces a
triangle of $G$.  If this is at the end of Step 1, this means that, up
to a permutation of $\{1, 2, 3\}$, the algorithm has found a vertex
$c_1\in X_1$ that has a neighbour $c_2\in X_2$ and a neighbour $c_3\in
X_3$.  Then the six vertices $b_1, b_2, b_3, c_1, c_2, c_3$ induce a
pyramid if $c_2, c_3$ are not adjacent or a prism if $c_2, c_3$ are
adjacent; so the positive answer is correct.  Now suppose that the
positive answer is returned at the end of step 3.  This means that
some component $H$ of $X$ gets the three labels $1, 2, 3$.  So, for
each $i=1, 2, 3$, the set $V_i$ of vertices of $H$ that have a
neighbour in $X_i$ is not empty.  We can apply
Lemma~\ref{lem:3sorties} to $H$, with the same notation, and we
consider the subgraph $F$ of $H$ described in the lemma, which leads
to the following three cases.  In each case we will see that $G$
contains a prism or a pyramid.

{\it Outcome~\ref{fp} of  Lemma~\ref{lem:3sorties}: $F$ is a chordless
path such that, up to a permutation of $V_1, V_2, V_3$, one end of $F$
is  a vertex  $v_1\in V_1$,  the other  is a  vertex $v_3\in  V_3$, no
interior vertex  of $F$ is in $V_1\cup  V_3$, and $F$ has  a vertex of
$V_2$.} There exists a neighbour  $c_1$ of $v_1$ in $X_1$, a neighbour
$c_3$  of $v_3$  in $X_3$,  and a  vertex $c_2$  of $X_2$  that  has a
neighbour in  $F$.  Note that  there is at  most one edge  among $c_1,
c_2, c_3$, for  otherwise we would have stopped at  Step 1.  Let $x,y$
be the neighbours of $c_2$  along $F$ that are closest respectively to
$v_1$  and $v_3$.   If $c_1,  c_2$ are  adjacent and  $y\not=v_1$ then
$c_2$-$c_1$-$b_1$, $c_2$-$b_2$, $c_2$-$y$-$F$-$v_3$-$c_3$-$b_3$ form a
pyramid,  while   if  $c_1,  c_2$   are  adjacent  and   $y=v_1$  then
$c_1$-$b_1$,  $c_2$-$b_2$, $v_1$-$F$-$v_3$-$c_3$-$b_3$  form  a prism.
So suppose $c_2$  is not adjacent to $c_1$ and  likewise not to $c_3$.
If   $c_1,  c_3$   are  adjacent   and  $v_1=v_3$   then  $c_1$-$b_1$,
$v_1$-$c_2$-$b_2$,  $c_3$-$b_3$  form  a  prism.  If  $c_1,  c_3$  are
adjacent and $v_1\not=v_3$ then either  $x\neq v_3$ or $y\neq v_1$, so
let  us assume  up to  symmetry  that $x\neq  v_3$; then  $c_1$-$b_1$,
$c_1$-$v_1$-$F$-$x$-$c_2$-$b_2$, $c_1$-$c_3$-$b_3$ form a pyramid.  So
suppose   $c_1,    c_3$   are    not   adjacent.    If    $x=y$   then
$x$-$F$-$v_1$-$c_1$-$b_1$,  $x$-$c_2$-$b_2$, $x$-$F$-$v_3$-$c_3$-$b_3$
form  a  pyramid.  If  $x,y$  are  different  and not  adjacent,  then
$c_2$-$x$-$F$-$v_1$-$c_1$-$b_1$,                           $c_2$-$b_2$,
$c_2$-$y$-$F$-$v_3$-$c_3$-$b_3$   form  a   pyramid.   If   $x,y$  are
different  and adjacent, then  $x$-$F$-$v_1$-$c_1$-$b_1$, $c_2$-$b_2$,
$y$-$F$-$v_3$-$c_3$-$b_3$ form a prism.

{\it  Outcome~\ref{fc})  of  Lemma~\ref{lem:3sorties}, with  the  same
notation.} For $i=1, 2, 3$, there exists a neighbour $c_i$ of $v_i$ in
$X_i$.  Since the vertices $v_1, v_2, v_3$ are pairwise different, for
each $i=1,  2, 3$,  vertex $c_i$  has no other  neighbour in  $F$ than
$v_i$.     If   $c_1,   c_2$    are   adjacent,    then   $c_1$-$b_1$,
$c_1$-$c_2$-$b_2$,  $c_1$-$v_1$-$F_1$-$f$-$F_3$-$v_3$-$c_3$-$b_3$ form
a pyramid.  So suppose, by symmetry, that $c_1, c_2, c_3$ are pairwise
not    adjacent.     Then    for     $i=1,    2,    3$    the    paths
$f$-$F_i$-$v_i$-$c_i$-$b_i$ form a pyramid.

{\it  Outcome~\ref{ft})  of  Lemma~\ref{lem:3sorties}, with  the  same
notation.} For $i=1, 2, 3$, there exists a neighbour $c_i$ of $v_i$ in
$X_i$.  Since the vertices $v_1, v_2, v_3$ are pairwise different, for
each  $i=1, 2,  3$ vertex  $c_i$ has  no other  neighbour in  $F$ than
$v_i$.     If   $c_1,   c_2$    are   adjacent,    then   $c_1$-$b_1$,
$c_1$-$c_2$-$b_2$,
$c_1$-$v_1$-$F_1$-$w_1$-$w_3$-$F_3$-$v_3$-$c_3$-$b_3$  form a pyramid.
So  suppose,  by symmetry,  that  $c_1,  c_2,  c_3$ are  pairwise  non
adjacent.      Then     for     $i=1,     2,     3$,     the     paths
$w_i$-$F_i$-$v_i$-$c_i$-$b_i$  form a  prism.  So  in either  case $G$
contains  a  pyramid or  a  prism, and  the  proof  of correctness  is
complete.

\emph{Complexity analysis:} Finding all triples takes time
$O(|V(G)|^3)$.  For each triple, computing the sets $X_1, X_2, X_3, X$
takes time $O(|V(G)|)$.  Finding the components of $X$ takes time
$O(|V(G)|^2)$.  Marking the components can be done as follows: for
each edge $uv$ of $G$, if $u$ is in a component $H$ of $X$ and $v$ is
in some $X_i$ then mark $H$ with label $i$; so this takes time
$O(|V(G)|^2)$.  Thus the overall complexity is $O(|V(G)|^5)$.  $\Box$

We observe that the above two algorithms are faster than the algorithm
from \cite{CS2002} for finding a pyramid.


\section{Recognition of graphs in class ${\cal A}$}
\label{sec:classa}

We can now present the algorithm for recognizing graphs in the class
$\cal A$.

\begin{algorithm}{Recognition of graphs in class $\cal A$}
\label{alg:classa}

\emph{Input:} A graph $G$.

\emph{Output:} The positive answer ``$G$  is in class $\cal A$'' if it
is; else the negative answer ``$G$ is not in class $\cal A$''.

\emph{Method:}\\ Step 1.  Test whether $G$ contains no antihole of
length at least $5$ as explained at the end of the introduction.\\
Step 2.  Test whether $G$ has no pyramid or prism using
Algorithm~\ref{alg:pyrpri2} above.\\ Step 3.  Test whether $G$ is
Berge using the algorithm from the preceding section.

\emph{Complexity:} $O(|V(G)|^9)$.

\end{algorithm}

The correctness of the algorithm is immediate from the correctness of
the algorithms it refers to and from the fact that Berge graphs
contain no pyramid.  The complexity is dominated by the last step of
the Berge recognition algorithm, which is $O(|V(G)|^9)$.  Note that
the other step of complexity $O(|V(G)|^9)$ in the Berge recognition
algorithm (deciding if the input graph contains a pyramid) can be
replaced by Step 2.  Additionally, we can remark that it is not
necessary to test for the existence of configurations of types T1,
\ldots, T4 when we call the Berge recognition algorithm,
because---this is not very hard to prove---any such configuration
contains an antihole of length at least $5$, so it is already excluded
by Step 2.  But this does not bring the overall complexity down from
$O(|V(G)|^9)$.

The algorithm for recognizing graphs in class $\cal A$ can also be
used to color graphs in class $\cal A$.  Recall that
Theorem~\ref{thm:maftro} states that: {\it If a graph $G$ is in class
$\cal A$ and is not a clique, it admits a pair of vertices whose
contraction yields a graph in class $\cal A$.} Therefore we could
enumerate all pairs of non-adjacent vertices of $G$ and test whether
their contraction produces a graph in class $\cal A$;
Theorem~\ref{thm:maftro} insures that at least one pair will work.  We
can then iterate this procedure until the contractions turn the graph
into a clique.  Since each vertex of the clique is the result of
contracting a stable set of $G$, a coloring of this clique corresponds
to an optimal coloring of $G$.  In terms of complexity, we may need to
check $O(|V(G)|^2)$ pairs at each contraction step, and there may be
$O(|V(G)|)$ steps.  So we end up with complexity $O(|V(G)|^{12})$.
This is not as good as the direct method from \cite{maftro02}, which
has complexity $O(|V(G)|^6)$.


\section{Even prisms}
\label{sec:evenprisms}

In this section we show how to decide in polynomial-time if a graph
that contains no odd hole contains an even prism.  Let $K$ be an even
prism, formed by paths $P_1, P_2, P_3$, with triangles $\{a_1, a_2,
a_3\}$ and $\{b_1, b_2, b_3\}$ so that for $1\le i\le 3$ path $P_i$ is
from $a_i$ to $b_i$.  Let $m_i$ be the middle vertex of path $P_i$.
We say that the $9$-tuple $(a_1, a_2, a_3, b_1, b_2, b_3, m_1, m_2,
m_3)$ is the \emph{frame} of $K$.  When we talk about a prism, the
word small refers to its number of vertices.

\begin{lemma}
\label{lem:epri}
Let $G$ be a graph that contains no odd hole and contains an even
prism, and let $K$ be a smallest even prism in $G$.  Let $K$ be formed
by paths $P_1, P_2, P_3$ and have frame $(a_1, a_2, a_3, b_1, b_2,
b_3, m_1, m_2, m_3)$, with $a_i, m_i, b_i\in V(P_i)$ ($1\le i\le 3$).
Let $R$ be any path of $G$ whose ends are $a_1, m_1$, whose interior
vertices are not adjacent to $a_2, a_3, b_2$ or $b_3$, and which is
shortest with these properties.  Then $a_1$-$R$-$m_1$-$P_1$-$b_1$ is a
chordless path $R_1$ and $R_1, P_2, P_3$ form a smallest even prism in
$G$.
\end{lemma}
\emph{Proof.} Let $k$ be the length (number of edges) of path $P_1$;
so $k$ is even.  Note that $|E(R)|\le k/2$ since the path
$a_1$-$P_1$-$m_1$ satisfies the properties required for $R$.  Call $Q$
the chordless path induced by $V(P_2)\cup V(P_3) \setminus \{a_2,
a_3\}$ and call $a'_2, a'_3$ the ends of $Q$ so that for $j=2, 3$
vertex $a'_j$ is adjacent to $a_j$.

Suppose that no interior vertex of $R$ has any neighbour in $Q$.  Let
$R_1$ be a shortest path from $a_1$ to $b_1$ contained in
$a_1$-$R$-$m_1$-$P_1$-$b_1$.  So $|E(R_1)|\le k$ and $R_1, P_2, P_3$
form a prism $K'$ with $|V(K')|\le |V(K)|$.  Since $G$ contains no odd
hole, $R_1$ has even length (else $V(R_1)\cup V(P_2)$ would induce an
odd hole), so $K'$ is an even prism.  Thus $K'$ is a smallest even
prism, and we have equality in the above inequalities; in particular
$R_1$ is equal to $a_1$-$R$-$m_1$-$P_1$-$b_1$ and the theorem holds.

We may now assume that some vertex $c$ of $R$ has a neighbour in $Q$,
and we choose $c$ closest to $m_1$ along $R$.  Let $S$ be a chordless
path from $c$ to $b_1$ contained in $c$-$R$-$m_1$-$P_1$-$b_1$.  We
have $|E(S)|< k$ since $|E(R)|\le k/2$ and $c\neq a_1$.  By the choice
of $c$ no vertex of $S\setminus b_1$ has a neighbour in $P_2$ or
$P_3$.  Let $x,y$ be the neighbours of $c$ along $Q$ that are closest
respectively to $a'_2$ and to $a'_3$.  If $x=y$ then $V(S)\cup V(P_2)
\cup V(P_3)$ induces a pyramid with triangle $\{b_1, b_2, b_3\}$ and
apex $x$, so $G$ contains an odd hole, a contradiction.  Thus $x\neq
y$.  If $x,y$ are not adjacent then $V(S) \cup V(P_2)\cup V(P_3)$
contains a pyramid with triangle $\{b_1, b_2, b_3\}$ and apex $c$, a
contradiction.  So $x,y$ are different and adjacent and, up to
symmetry, we may assume that they lie in the interior of $P_2$.  Now
$V(S)\cup V(P_2)\cup V(P_3)$ induces a prism $K'$, with triangles
$\{b_1, b_2, b_3\}$ and $\{c, x, y\}$, and $|V(K')|< |V(K)|$ since
$|E(S)|< k$.  Thus $K'$ is an odd prism, which means that
$y$-$P_2$-$b_2$ is an odd path, and so $a_2$-$P_2$-$x$ is an even
path.  Let $R'$ be a chordless path from $c$ to $a_1$ contained in
$c$-$R$-$m_1$-$P_1$-$a_1$.  We have $|E(R')|< k$ since $|E(R)|\le
k/2$.  By the choice of $c$ no vertex of $R'\setminus a_1$ has a
neighbour in $P_2$ or $P_3$.  Then $R'$ has even length for otherwise
$V(R')\cup V(a_2$-$P_2$-$x)$ induces an odd hole.  Now $V(R')\cup
V(P_2)\cup V(P_3)$ induces a prism $K''$ with triangles $\{a_1, a_2,
a_3\}$ and $\{c, x, y\}$, and $K''$ is an even prism, and we have
$|V(K'')|< |V(K)|$ since $|E(R')|< k$.  This is a contradiction, which
completes the proof.  $\Box$

Now we can give an algorithm:

\begin{algorithm}{Detection of an even prism in a graph that contains
no odd hole}
\label{alg:epri}

\emph{Input:} A graph $G$ that contains no odd hole.

\emph{Output:} An induced even prism of $G$ if $G$ contains any; else
the negative answer ``$G$ does not contain an even prism.''

\emph{Method:} For every $9$-tuple $(a_1, a_2, a_3, b_1, b_2, b_3,
m_1, m_2, m_3)$ of vertices of $G$ such that $\{a_1, a_2, a_3\}$ and
$\{b_1, b_2, b_3\}$ induce triangles, do: \\ For $i=1, 2, 3$, compute
the set $F_i$ of those vertices that are not adjacent to $a_{i+1},
a_{i+2}, b_{i+1}, b_{i+2}\}$ (with indices modulo $3$); look for a
shortest path $R_i$ from $a_i$ to $m_i$ whose interior vertices are in
$F_i$, and look for a shortest path $S_i$ from $m_i$ to $b_i$ whose
interior vertices are in $F_i$.  If the six paths $R_1, R_2, R_3, S_1,
S_2, S_3$ exist and their vertices induce an even prism, then return
this prism and stop.  \\ If no $9$-tuple yields an even prism, return
the negative answer.

\emph{Complexity:} $O(|V(G)|^{11})$.
\end{algorithm}

\emph{Proof of correctness.} If the algorithm returns an even prism
then clearly $G$ contains this prism.  So suppose conversely that $G$
contains an even prism.  Let $K$ be a smallest even prism, and let
vertices $a_1, a_2, a_3,$ $b_1, b_2, b_3,$ $m_1, m_2, m_3$ be the
frame of $K$.  When the algorithm considers this $9$-tuple, it will
find paths $R_1, R_2, R_3, S_1, S_2, S_3$ since some paths in $K$ do
have the required properties.  Then, six applications of
Lemma~\ref{lem:epri} imply that the vertices of these six paths do
induce an even prism of $G$.  So the algorithm will detect this
subgraph.

Complexity analysis: Testing all $9$-tuples take time $O(|V(G)|^9)$.
For each $9$-tuple, finding the six paths takes time $O(|V(G)|^2)$ and
checking that the corresponding subgraph is an even prism takes time
$O(|V(G)|)$.  Thus the overall complexity is $O(|V(G)|^{11})$.  $\Box$


\section{Line-graphs of subdivisions of $K_4$}
\label{sec:lgbsk4}

The \emph{line-graph} of  a graph $R$ is the  graph whose vertices are
the  edges  of  $R$  and  where  two  vertices  are  adjacent  if  the
corresponding    edges    of   $R$    have    a   common    endvertex.
\emph{Subdividing} an  edge $xy$  in a graph  means replacing it  by a
path of length  at least two.  A \emph{subdivision} of  a graph $R$ is
any graph obtained by repeatedly subdividing edges.  Berge graphs that
do not contain the line-graph of a bipartite subdivision of $K_4$ play
an  important  role   in  the  proof  of  the   Strong  Perfect  Graph
Theorem~\cite{CRST2002}.  Thus recognizing them  may be of interest on
its own.  Moreover, solving this question is also useful for later use
in  the  recognition   of  graphs  in  the  class   ${\cal  A}'$  (see
Section~\ref{sec:aprime}).  Again it turns  out that decide if a graph
contains the  line-graph of a  subdivision of $K_4$ is  NP-complete in
general, see Section~\ref{sec:npc}.

\begin{figure}
\begin{center}
   \includegraphics{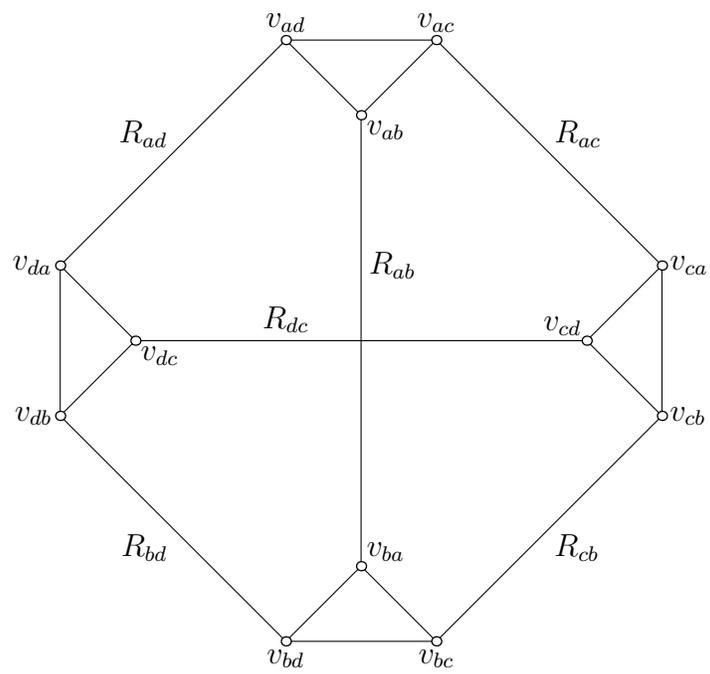}
\end{center}
\caption{{Line-graph of a subdivision of $K_4$}}\label{fig:FLK4}
\end{figure}

We will first deal with subdivisions of $K_4$ that are not necessarily
bipartite, but are not too trivial in the following sense: say that a
subdivision of $K_4$ is \emph{proper} if at least one edge of the
$K_4$ is subdivided.  It is easy to see that the line-graph of a
subdivision of $K_4$ is proper if and only if it has a vertex that
lies in only one triangle.  If $F$ is the line-graph of a proper
subdivision $R$ of $K_4$, let us denote by $a, b,c, d$ the four
vertices of $K_4$, i.e., the vertices of degree $3$ in $R$.  Then the
three edges incident to each vertex $x = a,b,c,d$ form a triangle in
$F$, which will be labelled $T_x$ and called a \emph{basic} triangle
of $F$.  ($F$ may have as many as two more, non-basic, triangles.)  In
$F$ there are six paths, each path being between vertices $x,y$ of
distinct triangles of $F$ (and so this path can be labelled $R_{xy}$
accordingly).  Note that $R_{xy}= R_{yx}$, and the six distinct paths
are vertex disjoint.  Some of these paths may have length $0$.  In the
basic triangle $T_x$, we denote by $v_{xy}$ the vertex that is the end
of the path $R_{xy}$.  Thus $F$ has paths $R_{ab}$, $R_{ac}$,
$R_{ad}$, $R_{bc}$, $R_{bd}$, $R_{cd}$, and the vertices of the basic
triangles of $F$ are $v_{ab}$, $v_{ac}$, $v_{ad}$, $v_{ba}$, $v_{bc}$,
$v_{bd}$, $v_{ca}$, $v_{cb}$, $v_{cd}$, $v_{da}$, $v_{db}$ and
$v_{dc}$.  The subgraph $F$ has no other edge than those in the four
basic triangles and those in the six paths.

For each of the six paths $R_{xy}$ of $F$, we call $m_{xy}$ one vertex
that is roughly in the middle of $R_{xy}$, so that if $\alpha$ denotes
the length of $v_{xy}$-$R_{xy}$-$m_{xy}$ and $\beta$ denotes the
length of $m_{xy}$-$R_{xy}$-$v_{yx}$, then $\alpha-\beta\in\{-1, 0,
1\}$.  Paths $R_{xy}$ are called the \emph{rungs} of $F$; vertices
$v_{xy}$ are called the \emph{corners} of $F$; and the $18$-tuple
$(v_{ab}, v_{ac}, \dots, v_{cd}, m_{ab}, \dots, m_{cd})$ is called a
\emph{frame} of $F$.

\begin{lemma}
\label{lem:lgpsk4}
Let $G$ be a graph that contains no pyramid.  Let $F$ be an induced
subgraph of $G$ that is the line-graph of a proper subdivision of
$K_4$ and $F$ has smallest size with this property, and let $(v_{ab},
v_{ac}, \dots, v_{cd}, m_{ab}, \dots, m_{cd})$ be a frame of $F$.  Let
$P$ be a path from $v_{ab}$ to $m_{ab}$ such that the interior
vertices of $P$ are non adjacent to every corner of $F$ other than
$v_{ab}$ and $P$ is a shortest path with these properties.  Then
$(V(F) \setminus V(R_{ab})) \cup V(P)$ induces the line-graph of a
proper subdivision of $K_4$ of smallest size.
\end{lemma}

\begin{figure}
\begin{center}
\includegraphics{fig.reco.3}
\end{center}
\caption{$F$ and $P$ for the proof of
Lemma~\protect\ref{lem:lgpsk4}}\label{fig:LK4}
\end{figure}

\emph{Proof.} Put $F' = F \setminus R_{ab}$.  If $v_{ab}, m_{ab}$ are
equal or adjacent, then $P=v_{ab}$-$R_{ab}$-$m_{ab}$ and the
conclusion is immediate.  So we may assume that $v_{ab}, m_{ab}$ are
distinct and not adjacent, which also implies $m_{ab} \neq v_{ba}$.

\begin{claim}
\label{clm:pnn}
If the  interior vertices of  $P$ have no  neighbour in $F'$  then the
lemma holds.
\end{claim}
\emph{Proof.} Let $u$ be  the vertex of $v_{ab}$-$P$-$m_{ab}$ that has
neighbours in  $m_{ab}$-$R_{ab}$-$v_{ba}$ and is  closest to $v_{ab}$.
Let $u'$ be the neighbour of $u$ in $m_{ab}$-$R_{ab}$-$v_{ba}$ closest
to   $v_{ba}$.   Then  $v_{ab}$-$P$-$u$-$u'$-$R_{ab}$-$v_{ba}$   is  a
chordless path  $R$, and $V(F')\cup  V(R)$ induce the line-graph  of a
proper subdivision of  $K_4$.  So this subgraph has  size at least the
size  of $F$,  which is  possible only  if $u=m_{ab}$,  and  this case
$V(F')\cup  V(R)$ induce  the line-graph  of a  proper  subdivision of
$K_4$ of smallest size, so the lemma holds.  $\Box$

Now we  may assume that there  exists a vertex $c_1\in  V(P)$ that has
neighbours in  $F'$, and choose  $c_1$ closest to $v_{ab}$  along $P$.
Also there exists  a vertex $d_1\in V(P)$ that  has neighbours in $F'$
and is  chosen closest to $m_{ab}$  along $P$.  Let us  show that this
leads  to a contradiction.   One may  look at  Figure~\ref{fig:LK4}.

\begin{claim}
\label{clm:nc1nd1}\ \\
1.  The set $N(c_1) \cap V(F')$ consists of an edge of $F'$.\\ 2.  The
set $N(d_1) \cap F'$ consists of an edge of $F'$.
\end{claim}
\emph{Proof.} Call  $H$ the hole induced by  $V(R_{ac}) \cup V(R_{bc})
\cup V(R_{bd}) \cup V(R_{ad})$.

First  suppose that  $c_1$  has no  neighbour  on $H$.   So $c_1$  has
neighbours  in  the interior  of  $R_{cd}$.   Let  $c_2, c_3$  be  the
neighbours of  $c_1$ respectively closest to $v_{cd}$  and to $v_{dc}$
along     $R_{cd}$.      If     $c_2=c_3$,     the     three     paths
$c_2$-$c_1$-$P$-$v_{ab}$,
$c_2$-$R_{cd}$-$v_{cd}$-$v_{ca}$-$R_{ca}$-$v_{ac}$,
$c_2$-$R_{cd}$-$v_{dc}$-$v_{da}$-$R_{ad}$-$v_{ad}$ form a pyramid with
triangle   $\{   v_{ab},  v_{ac},   v_{ad}\}$   and   apex  $c_2$,   a
contradiction.  If $c_2, c_3$ are distinct and not adjacent, the three
paths                                               $c_1$-$P$-$v_{ab}$,
$c_1$-$c_2$-$R_{cd}$-$v_{cd}$-$v_{ca}$-$R_{ca}$-$v_{ac}$,
$c_1$-$c_3$-$R_{cd}$-$v_{dc}$-$v_{da}$-$R_{ad}$-$v_{ad}$     form    a
pyramid with triangle $\{ v_{ab}, v_{ac}, v_{ad}\}$, and apex $c_1$, a
contradiction.   If $c_2,  c_3$ are  adjacent, we  have item~1  of the
claim.

Now suppose  that $c_1$ has  neighbours on $H$.  Define  two chordless
subpaths of $H$: $H_{ac} = H\setminus v_{ad}$ and $H_{ad} = H\setminus
v_{ac}$.  Let $c_2$  be the neighbour of $c_1$  on $H_{ac}$ closest to
$v_{ac}$, and let $c_3$ be  the neighbour of $c_1$ on $H_{ad}$ closest
to $v_{ad}$.  If $c_2=c_3$ then $V(H)\cup V(c_1$-$P$-$v_{ab})$ induces
a pyramid with triangle $\{  v_{ab}, v_{ac}, v_{ad}\}$ and apex $c_2$,
a contradiction.  So  $c_2 \neq c_3$.  If $c_2,  c_3$ are not adjacent
then         the        three         paths        $c_1$-$P$-$v_{ab}$,
$c_1$-$c_2$-$H_{ac}$-$v_{ac}$,  $c_1$-$c_3$-$H_{ad}$-$v_{ad}$  form  a
pyramid with triangle $\{ v_{ab},  v_{ac}, v_{ad}\}$ and apex $c_1$, a
contradiction.  So $c_2, c_3$ are adjacent and are the only neighbours
of $c_1$ on $H$.   Up to a symmetry, and by the  definition of $R$, we
may  assume  that  $c_2, c_3$  are  in  the  interior of  $R_{ac}$  or
$R_{bc}$.   If $c_1$ has  no neighbour  on $R_{cd}$  then conclusion~1
holds.  So  suppose that $c_1$ has  a neighbour $c_4$  on $R_{cd}$ and
$c_4$   is    closest   to    $v_{dc}$.    Then   the    three   paths
$c_1$-$P$-$v_{ab}$,
$c_1$-$c_4$-$R_{cd}$-$v_{dc}$-$v_{da}$-$R_{da}$-$v_{ad}$,
$c_1$-$c_2$-$H_{ac}$-$v_{ac}$ form a pyramid with triangle $\{ v_{ab},
v_{ac}, v_{ad}\}$ and apex  $c_1$, a contradiction.  This complete the
proof of item~1.

The  proof  of  item~2  is  similar, with  the  following  adjustment:
whenever path $c_1$-$P$-$v_{ab}$  was used for item~1, we  can use for
item~2  a   chordless  path  from  $d_1$  to   $v_{ba}$  contained  in
$d_1$-$P$-$m_{ab}$-$R_{ab}$-$v_{ba}$.  This completes the proof of the
claim.  $\Box$

\begin{claim}
\label{clm:j}
If $J$ is the line-graph of a subdivision of $K_4$ with $V(J)\subseteq
V(F')  \cup V(P)$  and  $c_1$ is  a corner  of  $J$, then  $J$ is  the
line-graph of a proper subdivision of $K_4$.
\end{claim}
\emph{Proof.} This claim follows  immediately from the fact that $c_1$
belongs to exactly one triangle of $J$.  $\Box$

In  view  of  Claim~\ref{clm:nc1nd1},   let  $c_2,  c_3$  be  the  two
neighbours of  $c_1$ in $F'$ and  $d_2, d_3$ be the  two neighbours of
$d_1$ in $F'$, with $c_2c_3, d_2d_3\in E(G)$.
\begin{claim}
\label{clm:c2c3}
We  may assume  that $c_2,  c_3$  lie in  $R_{ac}$ and  $d_2, d_3$  in
$R_{cb}$ or $R_{bd}$.
\end{claim}
\emph{Proof.} Recall from  the definition of $P$ that  $c_2, c_3, d_2,
d_3$ cannot be  corners of $F$.  If $c_2 c_3$ is  an edge of $R_{cd}$,
then   $V(v_{ab}$-$P$-$c_1)  \cup   V(R_{ac})   \cup  V(R_{ad})   \cup
V(R_{cd})$ induces the line-graph of  a subdivision de $K_4$, which is
proper  by  Claim~\ref{clm:j} and  is  strictly  smaller  than $F$,  a
contradiction.    If  $c_2  c_3$   is  an   edge  of   $R_{bc}$,  then
$V(v_{ab}$-$P$-$c_1)   \cup  V(F')$  induces   the  line-graph   of  a
subdivision  of $K_4$,  which is  proper by  Claim~\ref{clm:j}  and is
strictly smaller than  $F$, a contradiction.  So $c_2  c_3$ is an edge
of $R_{ac}$ or $R_{ad}$.  Similarly we may assume that $d_2 d_3$ is an
edge  of $R_{bc}$  or $R_{bd}$.   Then  by symmetry  the claim  holds.
$\Box$

We may assume that $v_{ac}, c_2, c_3, v_{ca}, d_2, d_3, v_{ad}$ appear
in this order along $H$.

\begin{claim}
\label{clm:c1d1}
Vertices $c_1, d_1$ are distinct and not adjacent.
\end{claim}
\emph{Proof.}  By Claims~\ref{clm:nc1nd1} and~\ref{clm:c2c3},  we know
that  $c_1,  d_1$  are  distinct.   If  they  are  adjacent,  the  set
$V(H')\cup \{c_1,  d_1\}$ induces the  line-graph of a  subdivision of
$K_4$, which  is proper by  Claim~\ref{clm:j} and is  strictly smaller
than $F$, a contradiction.  $\Box$

Let $e_1$ be the vertex of $c_1$-$P$-$v_{ab}$ that has a neighbour
$e_2$ in the interior of $m_{ab}$-$R_{ab}$-$v_{ab}$ and is closest to
$c_1$.  Let $e_4$ be the vertex of $d_1$-$P$-$m_{ab}$ that has
neighbour a neighbour $e_3$ in the interior of
$m_{ab}$-$R_{ab}$-$v_{ab}$, and is closest to $d_1$.  Given $e_1,
e_4$, take $e_2, e_3$ as close to each other as possible along
$R_{ab}$.

\begin{claim}
\label{clm:e1}
$e_1 \neq v_{ab}$.
\end{claim}
\emph{Proof.}  For   suppose  $e_1=v_{ab}$.   Then   the  three  paths
$v_{ab}$-$P$-$c_1$,                   $v_{ab}$-$v_{ac}$-$R_{ac}$-$c_2$,
$v_{ab}$-$R_{ab}$-$e_3$-$e_4$-$P$-$d_1$-$d_2$-$H_{ac}$-$c_3$   form  a
pyramid  with  triangle  $\{c_1,  c_2,  c_3\}$ and  apex  $v_{ab}$,  a
contradiction.  $\Box$

At       this       point       we      have       obtained       that
$c_1$-$P$-$e_1$-$e_2$-$R_{ab}$-$e_3$-$e_4$-$P$-$d_1$  is  a  chordless
path $R$ whose interior vertices  have no neighbour in $F'$.  Moreover
the subgraph $F_R$ induced by $V(F')\cup  V(R)$ is the line graph of a
subdivision of $K_4$, and it is proper by Claim~\ref{clm:j}.

\begin{claim}
\label{clm:fr}
$|V(F_R)|<|V(F)|$.
\end{claim}
\emph{Proof.} We need only show that  the total length of the rungs of
$F_R$ is strictly  smaller than the total length of  the rungs of $F$.
Let $\alpha$ be the  length of $v_{ab}$-$R_{ab}$-$m_{ab}$, let $\beta$
be the  length of $v_{ba}$-$R_{ab}$-$m_{ab}$, and let  $\delta$ be the
number of those edges of $F'$ that belong to the rungs of $F$.

The total length $l$ of the rungs of $F$ is equal to $\alpha + \beta +
\delta = 2 \alpha - \varepsilon + \delta$, with $\varepsilon = \alpha
- \beta \in \{-1, 0,1\}$ by the definition of $m_{ab}$.

The total length $l_R$ of the rungs of $F_R$ is at most $\delta
+2\alpha -3 $, and it is equal to this value only in the following
case: $e_4=m_{ab}$, there is only one vertex of $R_{ab}$ between $c_1$
and $d_1$, $e_1 v_{ab}\in E(G)$, $e_2 v_{ab}\in E(G)$, and the paths
$P$ and $v_{ab}$-$R_{ab}$-$m_{ab}$ have the same length.  Indeed in
this case the length of the rung of $F_R$ whose ends are $c_1, d_1$ is
equal to $2\alpha -3$.

Thus in either case we have $l_R < l$ and the claim holds.  $\Box$

Now the  preceding claim  leads to a  contradiction, which  proves the
lemma.  $\Box$

Lemma~\ref{lem:lgpsk4} is the basis of an algorithm for deciding if a
graph contains a pyramid or the line-graph of a proper subdivision of
$K_4$.

\begin{algorithm}{Detection of a line-graph of a proper subdivision of
$K_4$ in a graph that contains no pyramid}
\label{alg:pyrlgpsk4}

\emph{Input:} A graph $G$ that contains no pyramid.

\emph{Output:} An induced subgraph of $G$ that is the line-graph of a
proper subdivision of $K_4$ (if $G$ contains any); else the negative
answer ``$G$ does not contain the line-graph of a proper subdivision
of $K_4$''.

\emph{Method:} For every $18$-tuple of vertices $(v_{ab}$, $v_{ac}$,
$\dots$, $v_{cd}$, $m_{ab}$, $\dots$, $m_{cd})$, do the following:

For each $i,j\in\{a,  b, c, d\}$ with $i\neq j$,  find a shortest path
$S_{ij}$ from $v_{ij}$ to $m_{ij}$;

If  the subgraph induced  by the  union of  the twelve  paths $S_{ij}$
($i,j\in\{a,b,c,d\}$,  $i\neq  j$)  is  the  line-graph  of  a  proper
subdivision of $K_4$, return this subgraph and stop.

If no $18$-tuple has produced such a subgraph, return the negative
answer.

\emph{Complexity:} $O(|V(G)|^{20})$.
\end{algorithm}

\emph{Proof of correctness.} When the algorithm returns the line-graph
of a proper subdivision of $K_4$, clearly this answer is correct.

Conversely, suppose that $G$ contains the line-graph of a proper
subdivision of $K_4$.  Then $G$ has an induced subgraph $F$ that is
the line-graph of a proper subdivision of $K_4$ and has minimal size.

At  some step the  algorithm will  consider an  $18$-tuple $(v_{ab}$,
$v_{ac}$, $\dots$, $v_{cd}$, $m_{ab}$,  $\dots$, $m_{cd})$ which is a
frame of  $F$.  The algorithm will  find the paths  $S_{ij}$ since the
corresponding  paths of  $F$ do  have the  required  properties.  With
twelve  applications of  Lemma~\ref{lem:lgpsk4}, it  follows  that the
subgraph formed  by these twelve paths  is the line-graph  of a proper
subdivision of $K_4$ and is actually a smallest such subgraph.  So the
algorithm will detect this subgraph.

Complexity analysis:  There are $O(|V(G)|^{18})$ frames  to test.  For
each  such subset,  finding  the shortest  paths  $S_{ij}$ takes  time
$O(|V(G)|^2)$,  and  checking  that  the  subgraph they  form  is  the
line-graph of  a proper subdivision  of $K_4$ takes  time $O(|V(G)|)$.
Thus the algorithm finishes in time $O(|V(G)|^{20})$.  $\Box$

Let us now focus on finding line-graphs of \emph{bipartite}
subdivisions of $K_4$.

\begin{lemma}
\label{lem:fodd}
Let $R$ be a subdivision of $K_4$ and $F$ be the line-graph of $R$.
Then either $R=K_4$, or $F$ contains an odd hole, or $R$ is a
bipartite subdivision of $K_4$.
\end{lemma}
\emph{Proof.} Suppose $R \neq K_4$.  Call $a, b, c, d$ the four
vertices of the $K_4$ of which $R$ is a subdivision (i.e., the
vertices of degree $3$ in $R$), and for $i,j \in \{a, b, c, d\}$ with
$i\neq j$, call $C_{ij}$ the subdivision of edge $ij$.  Suppose that
$F$ contains no odd hole and $R$ is not bipartite.  Then $R$ contains
an odd cycle $Z$.  This cycle must be a triangle, for otherwise $L(R)$
contains an odd hole, a contradiction.  So me may assume up to
symmetry that $a, b, c$ induce a triangle.  Since $R\neq K_4$, we may
assume that $C_{ad}$ has length at least $2$.  But then one of
$E(C_{ad}) \cup \{ad\} \cup E(C_{cd})$ or $E(C_{ad}) \cup \{ab\} \cup
\{bc\} \cup C_{cd}$ is the edge set of an odd cycle of $R$, of length
at least $5$, so $L(R)$ contains an odd hole, a contradiction.  $\Box$

Now we can devise an algorithm that decides if a graph with no odd
hole contains the line-graph of a bipartite subdivision of $K_4$.
This algorithm is simply Algorithm~\ref{alg:pyrlgpsk4} applied to
graphs that contain no odd hole, by the preceding lemma.


\section{Recognition of graphs in class ${\cal A}'$}
\label{sec:aprime}

To decide if a graph is in class ${\cal A}'$, it suffices to decide
separately if it is Berge, if it has an antihole of length at least
$5$, and if it contains an odd prism.  But again it turns out that
this third question---deciding if a graph contains an odd prism---is
NP-complete (see Section~\ref{sec:npc}).  However, we can decide in
polynomial time if a graph with no odd hole contains an odd prism.
For this purpose the next lemmas will be useful.

\begin{lemma}
\label{lem:lgpsoddprism}
Let $F$ be  the line-graph of a bipartite  subdivision of $K_4$.  Then
$F$ contains an odd prism.
\end{lemma}
\emph{Proof.} Let $R$ be a bipartite subdivision of $K_4$ such that
$F$ is the line-graph of $R$, and let $a,b,c,d$ be the four vertices
of degree $3$ in $R$.  We may suppose without loss of generality that
$a, b$ lie on the same side of the bipartition of $R$.  Thus edge $ab$
is subdivided to a path $R_{ab}$ of even length, with the usual
notation.  Now it is easy to see that $F\setminus V(R_{cd})$ is an odd
prism.  $\Box$

\begin{figure}
\begin{center}
\includegraphics{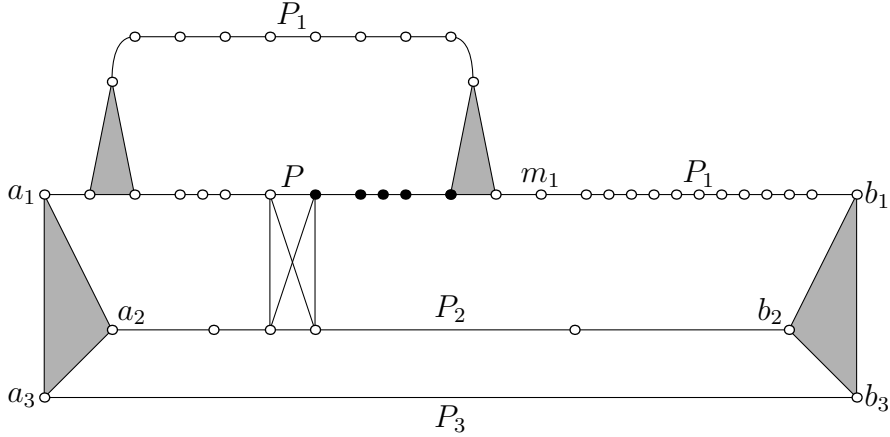}
\end{center}
\caption{A graph with six odd prisms}\label{fig:6LK4}
\end{figure}

Before we present an algorithm for recognizing graphs in class ${\cal
A}'$, we can remark that the technique which worked well for detecting
even prisms tends to fail for odd prisms.  The graph featured in
Figure~\ref{fig:6LK4} illustrates this problem.  This graph $G$ is the
line-graph of a bipartite graph, so it is a Berge graph.  For any two
grey triangles, there exists one (and only one) odd prism that contain
these two triangles.  Moreover, the paths $P_1, P_2, P_3$ form an odd
prism of $G$ of minimal size.  Yet, replacing $P_1$ (or the path
$a_1$-$P_1$-$m_1$) by a shortest path with the same ends does not
produce an odd prism.  Thus an algorithm that would be similar to the
even prism testing algorithm presented above may work incorrectly.  We
note however that in this example the graph $G$ contains the
line-graph of a proper subdivision of $K_4$ (the subgraph obtained by
forgetting the black vertices).  The next lemma shows that this remark
holds in general.

\begin{lemma}
\label{lem:oddprism}
Let $G$ be  a graph that contains  no odd hole and no  line-graph of a
proper  subdivision  of  $K_4$.  Let  $H$  be  a  prism in  $G$,  with
triangles $\{a_1, a_2, a_3\}$ and $\{b_1, b_2, b_3\}$.  Let $P$ be any
chordless path  from $a_1$  to $b_1$ whose  interior vertices  are not
adjacent to $a_2$, $a_3$, $b_2$,  $b_3$.  Then the three paths $P, P_2,
P_3$ form a prism of $G$ of the same parity as $H$.
\end{lemma}
\emph{Proof.} If the interior vertices of $P$ have no neighbour on
$P_2\cup P_3$ then the lemma holds.  So suppose that some interior
vertex $c_1$ of $P$ has neighbours on $P_2\cup P_3$, and choose $c_1$
closest to $a_1$ along $P$.  Define paths $H_2 = P_2 + P_3 \setminus
\{b_3\}$ and $H_3 = P_2 + P_3 \setminus \{b_3\}$.  For $i=2, 3$, let
$c_i$ be the neighbour of $c_1$ closest to $b_i$ along $H_i$.

If $c_2 = c_3$, then the three paths $c_2$-$c_1$-$P$-$a_1$,
$c_2$-$P_2$-$a_2$, $c_2$-$P_2$-$b_2$-$b_3$-$P_3$-$a_3$ form a pyramid
with triangle $\{a_1, a_2, a_3\}$ and apex $c_2$, a contradiction.  So
$c_2\neq c_3$.  If $c_2, c_3$ are not adjacent, then the three paths
$c_1$-$P$-$a_1$, $c_1$-$c_2$-$P_2$-$a_2$,
$c_1$-$c_3$-$P_2$-$b_2$-$b_3$-$P_3$-$a_3$ form a pyramid with triangle
$\{a_1, a_2, a_3\}$ and apex $c_1$, a contradiction.  So $c_2, c_3$
are adjacent.  Up to symmetry, $c_2 c_3$ is an edge of $P_2$.  If
$c_1, b_1$ are adjacent, then the three paths $c_1$-$b_1$,
$c_1$-$c_3$-$P_2$-$b_2$, $c_1$-$P$-$a_1$-$a_3$-$P_3$-$b_3$ form a
pyramid with triangle $\{b_1, b_2, b_3\}$ and apex $c_1$.  So we may
assume that $c_1, b_1$ are not adjacent.  Let $a'_1$ be the neighbour
of $a_1$ in $P_1$.  Let $d_1$ be the vertex of $a'_1$-$P$-$c_1$ that
has neighbours in $P_1$ and is closest to $c_1$.  Let $d_2, d_3$ be
the neighbours of $d_1$ along $P_1$ that are closest to $a_1$ and
$b_1$ respectively.

If  $d_2   =  d_3$,   then  the  three   paths  $d_2$-$d_1$-$P$-$c_1$,
$d_2$-$P_1$-$a_1$-$a_2$-$P_2$-$c_2$,
$d_2$-$P_1$-$b_1$-$b_2$-$P_2$-$c_3$  form   a  pyramid  with  triangle
$\{c_1,  c_2, c_3\}$  and apex  $d_2$, a  contradiction.   So $d_2\neq
d_3$.   If  $d_2,  d_3$  are   not  adjacent,  then  the  three  paths
$d_1$-$d_2$-$P_1$-$a_1$,     $d_1$-$d_3$-$P_1$-$b_1$-$b_3$-$P_3$-$a_3$,
$d_1$-$P$-$c_1$-$c_2$-$P_2$-$a_2$ form a pyramid with triangle $\{a_1,
a_2,  a_3\}$ and  apex  $d_1$,  a contradiction.   So  $d_2, d_3$  are
adjacent.  Then  the four triangles $\{a_1, a_2,  a_3\}$, $\{b_1, b_2,
b_3\}$, $\{c_1,  c_2, c_3\}$,  $\{d_1, d_2, d_3\}$  and the  six paths
$P_3$,    $a_2$-$P_2$-$c_2$,   $a_1$-$P_1$-$d_2$,   $b_2$-$P_2$-$c_3$,
$b_1$-$P_1$-$d_3$,   $c_1$-$P$-$d_1$   form   the  line-graph   of   a
subdivision  of $K_4$, and  it is  not the  line-graph of  $K_4$ since
$a_3\neq b_3$; so $G$ contains  the line-graph of a proper subdivision
of $K_4$, a contradiction.  $\Box$

Now we can present an algorithm that decides if a graph with no odd
hole contains an odd prism.

\begin{algorithm}{Detection of an odd prism in a graph that contains
no odd hole}
\label{alg:opri}

\emph{Input:} A graph $G$ that contains no odd hole.\\
\emph{Output:} An odd prism induced in $G$, if $G$ contains any, else
the negative answer ``$G$ contains no odd prism''.

\emph{Method:} Using Algorithm~\ref{alg:pyrlgpsk4}, test whether $G$
contains the line-graph of a proper subdivision of $K_4$.  If $G$
contains such a subgraph $F$, for each of the six rungs $R$ of $F$,
test if $F\setminus V(R)$ is an odd prism, and if it is, return this
odd prism.  If Algorithm~\ref{alg:pyrlgpsk4} answers that $G$ does not
contain the line-graph of a proper subdivision of $K_4$, then for
every $6$-tuple $(a_1, a_2, a_3, b_1, b_2, b_3)$ do:

For $i=1,  2, 3$  compute a  shortest path $P_i$  from $a_i$  to $b_i$
whose  interior vertices  are  not adjacent  to $a_{i+1}$,  $a_{i+2}$,
$b_{i+1}$  and $b_{i+2}$  (subscripts are  understood modulo  $3$).  If
paths $P_1, P_2,  P_3$ exist and form an odd  prism, return the answer
no and stop.

If no $6$-tuple has produced a prism, return the answer yes.

\emph{Complexity:} $O(|V(G)|^{20})$.

\end{algorithm}

\emph{Proof of correctness.} If $G$ contains the line-graph of proper
subdivision of $K_4$, this will be detected by
Algorithm~\ref{alg:pyrlgpsk4}.  If $G$ contains no odd hole and no odd
prism, then Lemma~\ref{lem:lgpsoddprism} ensures that $G$ cannot
contain the line-graph of a proper subdivision of $K_4$.  So the
algorithm will return the correct answer.

Now suppose that $G$ does not contain the line graph of a proper
subdivision of $K_4$ and $G$ contains an odd prism, with triangles
$\{a_1, a_2, a_3\}$ and $\{b_1, b_2, b_3\}$.  Then in some step the
algorithm will consider these six vertices, and it will find paths
$P_i$ since the corresponding paths of the prism have the required
properties.  By three applications of Lemma~\ref{lem:oddprism}, we
obtain that $P_1, P_2, P_3$ form an odd prism, and so the algorithm
will detect it.

\emph{Complexity  analysis:} The complexity  is clearly  determined by
its costliest step, which is Algorithm~4.  $\Box$

Now deciding if a graph is in class ${\cal A}'$ can be done as
follows: test if $G$ contains an antihole of length at least $5$ as
explained earlier; test if $G$ is Berge using the algorithm from
Section~\ref{sec:berge}; then use Algorithm~\ref{alg:opri} to test if
$G$ contains no odd prism.  The complexity is the same as that of
Algorithm~\ref{alg:opri}.

We note that if Conjecture~\ref{conj:pc} is true then the algorithm
for recognizing graphs in class ${\cal A}'$ can be used to color
optimally the vertices of any graph $G\in {\cal A}'$ (even if a proof
of Conjecture~\ref{conj:pc} is not algorithmic); this can be done
similarly to the remark made at the end of Section~\ref{sec:classa},
as follows.  Enumerate all pairs of non-adjacent vertices of $G$ and
test whether their contraction produces a graph in class ${\cal A}$;
the assumed validity of Conjecture~\ref{conj:pc} insures that at least
one pair will work.  Then iterate this procedure until the
contractions turn the graph into a clique.  In terms of complexity,
since we may need to check $O(|V(G)|^2)$ pairs at each contraction
step, and there may be $O(|V(G)|)$ steps, we end up with total
complexity $O(|V(G)|^{23})$; thus it is desirable to find a proof of
Conjecture~\ref{conj:pc} that produces an algorithm with lower
complexity.


\section{NP-complete problems}
\label{sec:npc}

In this section we show that the following problems are NP-complete:
\begin{itemize}
\item
Decide if a graph contains a prism.
\item
Decide if a graph contains an even prism.
\item
Decide if a graph contains an odd prism.
\item
Decide if a  graph contains the line-graph of  a proper subdivision of
$K_4$.
\item
Decide if a graph contains the line-graph of a bipartite subdivision
of $K_4$.
\end{itemize}
We have seen in the preceding sections that all these problems are
polynomial when the input is restricted to the class of graphs that
contain no odd hole.

The above NP-completeness results can all be derived from the
following theorem.  Let us call problem $\Pi$ the decision problem
whose input is a triangle-free graph $G$ and two non-adjacent vertices
$a,b$ of $G$ of degree $2$ and whose question is: ``Does $G$ have a
hole that contains both $a,b$?''  Bienstock \cite{bien} mentions that
this problem is NP-complete in general (i.e., not restricted to
triangle-free graphs).  We adapt his proof here for triangle-free
graphs.

\begin{theorem}
\label{thm:pinpc}
Problem $\Pi$ is NP-complete.
\end{theorem}
\emph{Proof.} Let us give a polynomial reduction from the problem {\sc
$3$-Satisfiability} of Boolean functions to problem $\Pi$.  Recall
that a Boolean function with $n$ variables is a mapping $f$ from $\{0,
1\}^n$ to $\{0, 1\}$.  A Boolean vector $\xi\in\{0, 1\}^n$ is a
\emph{truth assignment} for $f$ if $f(\xi)=1$.  For any Boolean
variable $x$ on $\{0, 1\}$, we write $\overline{x}:=1-x$, and each of
$x, \overline{x}$ is called a \emph{literal}.  An instance of {\sc
$3$-Satisfiability} is a Boolean function $f$ given as a product of
clauses, each clause being the Boolean sum $\vee$ of three literals;
the question is whether $f$ admits a truth assignment.  The
NP-completeness of {\sc $3$-Satisfiability} is a fundamental result in
complexity theory, see \cite{garjoh79}.

Let $f$ be an instance  of {\sc $3$-Satisfiability}, consisting of $m$
clauses $C_1, \ldots,  C_m$ on $n$ variables $x_1,  \ldots, x_n$.  Let
us build a graph $G_f$  with two specialized vertices $a,b$, such that
there will be a hole containing both $a,b$ in $G$ if and only if there
exists a truth assignment for $f$.

\begin{figure}[p]
\begin{center}
\includegraphics{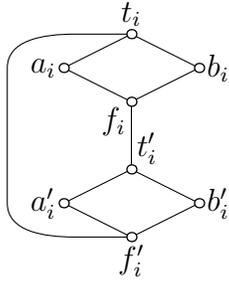}
\end{center}
\caption{Graph $G(x_i)$}\label{fig:gxi}
\end{figure}

For each variable $x_i$ ($i=1, \ldots, n$), make a graph $G(x_i)$ with
eight vertices $a_i,  b_i, t_i, f_i, a'_i, b'_i,  t'_i, f'_i,$ and ten
edges  $a_it_i, a_if_i,  b_it_i, b_if_i$  (so that  $\{a_i,  b_i, t_i,
f_i\}$ induces  a hole), $a'_it'_i, a'_if'_i,  b'_it'_i, b'_if'_i$ (so
that  $\{a'_i,  b'_i, t'_i,  f'_i\}$  induces  a  hole) and  $t_if'_i,
t'_if_i$.  See Figure~\ref{fig:gxi}.

\begin{figure}[p]
\begin{center}
\includegraphics{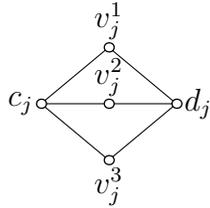}
\end{center}
\caption{Graph $G(C_j)$}\label{fig:gcj}
\end{figure}

\begin{figure}[p]
\begin{center}
\includegraphics{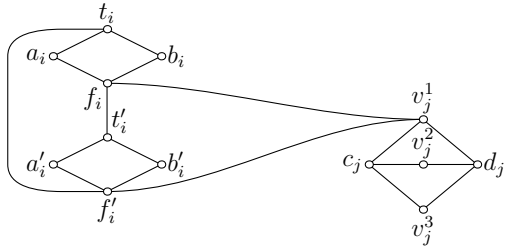}
\end{center}
\caption{The two edges added to $G_f$ in the case $u_j^p=x_i$}\label{fig:gf}
\end{figure}

\begin{figure}[p]
\begin{center}
\includegraphics{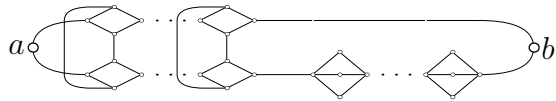}
\end{center}
\caption{Graph $G_f$}\label{fig:gf2}
\end{figure}

For  each  clause  $C_j$   ($j=1,  \ldots,  m$),  with  $C_j=u_j^1\vee
u_j^2\vee u_j^3$, where  each $u_j^p$ ($p=1, 2, 3$)  is a literal from
$\{x_1, \ldots, x_n, \overline{x}_1, \ldots, \overline{x}_n\}$, make a
graph $G(C_j)$ with five vertices  $c_j, d_j, v_j^1, v_j^2, v_j^3$ and
six edges  so that each of $c_j,  d_j$ is adjacent to  each of $v_j^1,
v_j^2,  v_j^3$.   See  Figure~\ref{fig:gcj}.   For  $p=1,  2,  3$,  if
$u_j^p=x_i$  then  add  two  edges  $u_j^pf_i,  u_j^pf'_i$,  while  if
$u_j^p=\overline{x}_i$ then add two edges $u_j^pt_i, u_j^pt'_i$.

The graph $G_f$ is obtained from the disjoint union of the $G(x_i)$'s
and the $G(C_j)$'s as follows.  For $i=1, \ldots, n-1$, add edges
$b_ia_{i+1}$ and $b'_ia'_{i+1}$.  Add an edge $b'_nc_1$.  For $j=1,
\ldots, m-1$, add an edge $d_jc_{j+1}$.  Introduce the two specialized
vertices $a,b$ and add edges $aa_1, aa'_1$ and $bd_m, bb_n$.  See
Figure~\ref{fig:gf}.  Clearly the size of $G_f$ is polynomial
(actually linear) in the size $n+m$ of $f$.  Moreover, it is easy to
see that $G_f$ contains no triangle, and that $a,b$ are non-adjacent
and both have degree $2$.

Suppose  that $f$ admits  a truth  assignment $\xi\in\{0,  1\}^n$.  We
build a hole  in $G$ by selecting vertices  as follows.  Select $a,b$.
For  $i=1, \ldots,  n$, select  $a_i, b_i,  a'_i, b'_i$;  moreover, if
$\xi_i=1$ select  $t_i, t'_i$, while if $\xi_i=0$  select $f_i, f'_i$.
For $j=1,  \ldots, m$, since $\xi$  is a truth assignment  for $f$, at
least  one  of the  three  literals  of $C_j$  is  equal  to $1$,  say
$u_j^p=1$  for some  $p\in\{1, 2,  3\}$.  Then  select $c_j,  d_j$ and
$v_j^p$.   Now it  is  a routine  matter  to check  that the  selected
vertices  induce a  cycle $Z$  that contains  $a,b$, and  that  $Z$ is
chordless, so it is a hole.  The  main point is that there is no chord
in $Z$ between some subgraph  $G(C_j)$ and some subgraph $G(x_i)$, for
that  would  be  either  an  edge  $t_iv_j^p$  (or  $t'_iv_j^p$)  with
$u_j^p=x_i$ and  $\xi_i=1$, or, symmetrically, an  edge $f_iv_j^p$ (or
$f'_iv_j^p$) with $u_j^p=\overline{x}_i$ and $\xi_i=0$, in either case
a contradiction to the way the vertices of $Z$ were selected.

Conversely, suppose that $G_f$ admits  a hole $Z$ that contains $a,b$.
Clearly $Z$ contains  $a_1, a'_1$ since these are  the only neighbours
of $a$ in $G_f$.

\begin{claim}\label{clm:zgxi}
For $i=1, \ldots, n$, $Z$ contains exactly six vertices of $G_i$: four
of them are $a_i, a'_i, b_i, b'_i$, and the other two are either $t_i,
t'_i$ or $f_i, f'_i$.
\end{claim}
\emph{Proof.} First we prove the claim for $i=1$.  Since $a, a_1$ are
in $Z$ and $a_1$ has only three neighbours $a, t_1, f_1$, exactly one
of $t_1, f_1$ is in $Z$.  Likewise exactly one of $t'_1, f'_1$ is in
$Z$.  If $t_1, f'_1$ are in $Z$ then the vertices $a, a_1, a'_1, t_1,
f'_1$ are all in $Z$ and they induce a hole that does not contain $b$,
a contradiction.  Likewise we do not have both $t'_1, f_1$ in $Z$.
Therefore, up to symmetry we may assume that $t_1, t'_1$ are in $Z$
and $f_1, f'_1$ are not.  If a vertex $u_j^p$ of some $G(C_j)$ ($1\le
j\le m$, $1\le p\le 3$) is in $Z$ and is adjacent to $t_1$ then, since
this $u_j^p$ is also adjacent to $t'_1$, we see that the vertices $a,
a_1, a'_1, t_1, t'_1, u_j^p$ are all in $Z$ and induce a hole that
does not contain $b$, a contradiction.  Thus the neighbour of $t_1$ in
$Z\setminus a_1$ is not in any $G(C_j)$ ($1\le j\le m$), so that
neighbour is $b_1$.  Likewise $b'_1$ is in $Z$.  So the claim holds
for $i=1$.  Since $b_1$ is in $Z$ and exactly one of $t_1, f_1$ is in
$Z$, and $b_1$ has degree $3$ in $G_f$, we obtain that $a_2$ is in
$Z$, and similarly $b_2$ is in $Z$.  Now the proof of the claim for
$i=2$ is essentially the same as for $i=1$, and by induction the claim
holds up to $i=n$.  $\Box$

\begin{claim}\label{clm:zgcj}
For  $j=1, \ldots,  m$, $Z$  contains $c_j,  d_j$ and  exactly  one of
$v_j^1, v_j^2, v_j^3$.
\end{claim}
\emph{Proof.} First we prove this claim for $j=1$.  By
Claim~\ref{clm:zgxi}, $b'_n$ is in $Z$ and exactly one of $t'_n, f'_n$
is in $Z$, so (since $b'_n$ has degree $3$ in $G_f$) $c_1$ is in $Z$.
Consequently exactly one of $u_1^1, u_1^2, u_1^3$ is in $Z$, say
$u_1^1$.  The neighbour of $u_1^1$ in $Z\setminus c_1$ cannot be a
vertex of some $G(x_i)$ ($1\le i\le n$), for that would be either
$t_i$ (or $f_i$) and thus, by Claim~\ref{clm:zgxi}, $t'_i$ (or $f'_i$)
would be a third neighbour of $u_1^1$ in $Z$, a contradiction.  Thus
the other neighbour of $u_1^1$ in $Z$ is $d_1$, and the claim holds
for $j=1$.  Since $d_1$ has degree $4$ in $G_f$ and exactly one of
$v_1^1, v_1^2, v_1^3$ is in $Z$, it follows that its fourth neighbour
$c_2$ is in $Z$.  Now the proof of the claim for $j=2$ is the same as
for $j=1$, and by induction the claim holds up to $j=m$.  $\Box$

We can now make a Boolean  vector $\xi$ as follows.  For $i=1, \ldots,
n$, if $Z$ contains $t_i, t'_i$ set $\xi_i = 1$; if $Z$ contains $f_i,
f'_i$ set  $\xi_i = 0$.   By Claim~\ref{clm:zgxi} this  is consistent.
Consider any  clause $C_j$  ($1\le j\le m$).   By Claim~\ref{clm:zgcj}
and up to symmetry we may assume  that $v_j^1$ is in $Z$.  If $u_j^1 =
x_i$  for some  $i\in\{1, ..,  n\}$,  then the  construction of  $G_f$
implies that $f_i, f'_i$ are not in $Z$, so $t_i, t'_i$ are in $Z$, so
$\xi_i=1$,  so  clause $C_j$  is  satisfied  by  $x_i$.  If  $u_j^1  =
\overline{x}_i$ for some $i\in\{1,  .., n\}$, then the construction of
$G_f$ implies that  $t_i, t'_i$ are not in $Z$, so  $f_i, f'_i$ are in
$Z$, so  $\xi_i=0$, so clause $C_j$ is  satisfied by $\overline{x}_i$.
Thus $\xi$ is a truth assignment for $f$.  This completes the proof of
the theorem.  $\Box$

Now we can prove the main result of this section.
\begin{theorem}
\label{thm:prismsnpc}
The following problems are NP-complete:
\begin{enumerate}
\item
Decide if a graph contains a prism.
\item
Decide if a graph contains an odd prism.
\item
Decide if a graph contains an even prism.
\item
Decide if a graph contains the line-graph of a proper subdivision of
$K_4$.
\item
Decide if a graph contains the line-graph of a bipartite subdivision
of $K_4$.
\end{enumerate}
\end{theorem}
\emph{Proof.} For each of these five problems we show a reduction from
problem $\Pi$ to this problem.  So let $(G, a, b)$ be any instance of
problem $\Pi$, where $G$ is a triangle-free graph and $a, b$ are
non-adjacent vertices of $G$ of degree $2$.  Let us call $a', a''$ the
two neighbours of $a$ and $b', b''$ the two neighbours of $b$ in $G$.

\begin{figure}[!htb]
\begin{center}
\includegraphics{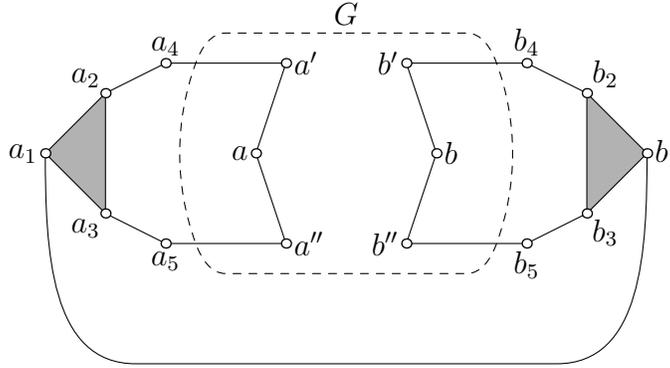}
\end{center}
\caption{Problem~1: $G$ and $G'$}
\label{fig:reco.8}
\end{figure}

\emph{Reduction to Problem~1:} Starting from $G$, build a graph $G'$
as follows (see Figure~\ref{fig:reco.8}): replace vertex $a$ by five
vertices $a_1, a_2, a_3, a_4, a_5$ with five edges $a_1 a_2$, $a_1
a_3$, $a_2 a_3$, $a_2 a_4$, $a_3 a_5$, and put edges $a_4 a'$ and $a_5
a''$.  Do the same with $b$, with five vertices named $b_1, \ldots,
b_5$ instead of $a_1, \ldots, a_5$ and with the analogous edges.  Add
an edge $a_1 b_1$.  Since $G$ has no triangle, $G'$ has exactly two
triangles $\{a_1, a_2, a_3\}$ and $\{b_1, b_2, b_3\}$.  Moreover we
see that $G'$ contains a prism if and only if $G$ contains a hole that
contains $a$ and $b$.  So every instance of $\Pi$ can be reduced
polynomially to an instance of Problem~1, which proves that Problem~1
is NP-complete.

\emph{Reduction to Problem 2:} Starting from $G$, build the same graph
$G'$ as above.  Then build eight graphs $G_{i,j,k}$ ($i, j, k \in
\{0,1\}$) as follows: if $i=1$, subdivide the edge $a_2 a_4$ into a
path of length $2$; else do not subdivide it.  Likewise, subdivide the
edge $a_3 a_5$ if and only if $j=1$; and subdivide the edge $a_1 b_1$
if and only if $k=1$.  Now $G$ contains a hole that contains $a$ and
$b$ if and only if at least one of the eight graphs $G_{i,j,k}$
contains an odd prism.  So every instance of $\Pi$ can be reduced
polynomially to eight instances of Problem 2.

\emph{Reduction to Problem 3:} Starting from $G$, build the eight
graphs $G_{i,j,k}$ as above.  Then $G$ contains a hole that contains
$a$ and $b$ if and only if at least one of the eight graphs
$G_{i,j,k}$ contains an even prism.  So every instance of $\Pi$ can be
reduced polynomially to eight instances of Problem 3.

\begin{figure}[!htb]
\begin{center}
\includegraphics{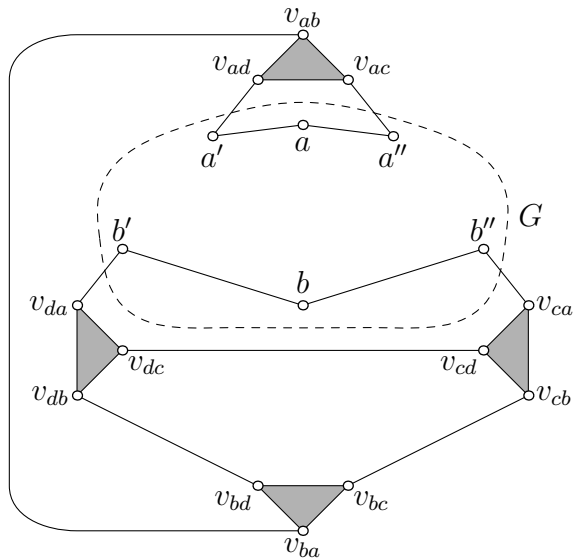}
\end{center}
\caption{Problem 4: $G$ and $G''$}
\label{fig:reco.9}
\end{figure}

\emph{Reduction to Problem 4:} Starting from $G$, build a graph $G''$
as follows (see Figure~\ref{fig:reco.9}): remove vertices $a$ and $b$
and add twelve vertices $v_{ab}$, $v_{ac}$, $v_{ad}$, $v_{ba}$,
$v_{bc}$, $v_{bd}$, $v_{ca}$, $v_{cb}$, $v_{cd}$, $v_{da}$, $v_{db}$,
$v_{dc}$.  Add edges such that each of $\{v_{ab}, v_{ac}, v_{ad}\}$,
$\{ v_{ba}, v_{bc}, v_{bd}\}$, $\{ v_{ca}, v_{cb}, v_{cd}\}$ and
$\{v_{da}, v_{db}, v_{dc}\}$ is a triangle.  Add edges $v_{ab}
v_{ba}$, $v_{dc} v_{cd}$, $v_{bd} v_{db}$, $v_{bc} v_{cb}$, $v_{ad}
a'$, $v_{ac} a''$, $v_{da} b'$, $v_{ca} b''$.  The graph $G''$
contains exactly four triangles, and $G$ contains a hole through $a$
and $b$ if and only if $G''$ contains the line-graph of a proper
subdivision of $K_4$.  So every instance of $\Pi$ can be reduced
polynomially to an instance of Problem 4.

\emph{Reduction to Problem 5:} Starting from $G''$, make four graphs
$G''_{i,j}$ ($i, j\in \{0,1\}$) as follows: if $i=1$ subdivide the
edge $v_{ad} a'$ into a path of length $2$, else do not subdivide it.
Subdivide likewise the edge $v_{ac} a''$ if and only if $j=1$.  Now
$G$ contains a hole through $a$ and $b$ if and only if one of the four
graphs $G''_{i,j}$ contains the line-graph of a bipartite subdivision
of $K_4$.  So every instance of $\Pi$ can be reduced polynomially to
four instances of Problem 5.  This completes the proof of the theorem.
$\Box$

\clearpage
\section{Conclusion}

We summarize the complexity results mentioned in this paper in the
following table, whose columns correspond to the class of graphs taken
as instances and whose lines correspond to the subgraph that we look
for.  The symbol $n$ refers to the number of vertices of the input
graph; 0 means trivial, NPC means NP-complete, and a question mark
means unsolved.

\vspace{4ex}

\begin{center}
\begin{tabular}{l|c|c|c|}
\  & General graphs &   Graphs with   &
  Graphs with  \\
  \ & & no pyramid & no odd hole \\ \hline
\rule{0ex}{3ex}Pyramid or prism & $n^5$ & $n^5$ & $n^5$  \\
Pyramid   & $n^9$ \cite{CS2002} & $0$ & $0$  \\
Prism & NPC & $n^5$ & $n^5$  \\
LGPS$K_4$ & NPC & $n^{20}$ & $n^{20}$  \\
LGBS$K_4$ & NPC & ? & $n^{20}$  \\
Odd prism & NPC & ? & $n^{20}$  \\
Even prism & NPC & ? & $n^{11}$  \\
  \hline
\end{tabular}
\end{center}



\clearpage

\begin{thebibliography}{99}

\bibitem{ber60} C.~Berge.  Les  probl\`emes de coloration en th\'eorie
des  graphes.   {\it Publ.   Inst.   Stat.   Univ.   Paris} 9  (1960),
123--160.

\bibitem{ber61}  C.~Berge.  F\"arbung  von Graphen,  deren s\"amtliche
bzw.~deren ungerade  Kreise starr sind  (Zusammenfassung).  {\it Wiss.
Z.  Martin Luther  Univ.  Math.-Natur.   Reihe}  (Halle-Wittenberg) 10
(1961), 114--115.

\bibitem{ber85}
C.~Berge.
\newblock {\em Graphs}.
\newblock North-Holland, Amsterdam/New York, 1985.

\bibitem{ber90}
M.E.~Bertschi, Perfectly contractile graphs.  {\it J. Comb.~Th.
B} {50} (1990), 222--230.

\bibitem{bien}
D. Bienstock.  On the complexity of testing for even holes and induced
odd paths.  \emph{Disc.~Math.} 90 (1991), 85--92.  Corrigendum in
\emph{Disc.~Math.} 102 (1992), 109.

\bibitem{CCLSV2002}
M.~Chudnovsky, G.~Cornu\'ejols, X.~Liu, P.~Seymour, K.~Vu\v{s}kovi\'c.
Cleaning for Bergeness.  Manuscript, 2002.

\bibitem{CRST2002}
M.~Chudnovsky, N.~Robertson, P.~Seymour, R.~Thomas.  The strong
perfect graph theorem.  Manuscript, Princeton Univ., 2002.

\bibitem{CS2002}
M.~Chudnovsky, P.~Seymour. Recognizing Berge graphs.  Manuscript,
Princeton Univ., 2002.

\bibitem{CLV2002}
G.~Cornu\'ejols, X.~Liu, K.~Vu\v{s}kovi\'c.  A polynomial algorithm
for recognizing perfect graphs.  Manuscript, Carnegie-Mellon Univ.,
2002.


\bibitem{epsbook}
H.~Everett, C.M.H.~de~Figueiredo, C.~Linhares~Sales, F.~Maffray,
O.~Porto, B.A.~Reed.  \newblock Even pairs.  In \cite{ramree01},
67--92.

\bibitem{fonuhr82}
J.~Fonlupt, J.P.~Uhry.  Transformations which preserve perfectness and
$h$-perfectness of graphs.  {\it Ann.~Disc.  Math.} {16} (1982),
83--85.

\bibitem{garjoh79}
M.R.~Garey, D.S.~Johnson.  {\it Computer and Intractability : A Guide
to the Theory of NP-completeness.} W.H.~Freeman, San Fransisco, 1979.

\bibitem{linmafree97}
C.~Linhares~Sales, F.~Maffray, B.A.~Reed.  On planar perfectly
contractile graphs.  {\it Graphs and Combin.} {13} (1997), 167--187.

\bibitem{maftro02}
F.~Maffray, N.~Trotignon.  A class of perfectly contractile graphs.
Research report 67, Laboratoire Leibniz, Grenoble, France,
http://www-leibniz.imag.fr/LesCahiers.  Submitted for publication.


\bibitem{ramree01}
J.L.~Ram\'{\i}rez-Alfons\'{\i}n, B.A.~Reed.  {\it Perfect Graphs}.
Wiley Interscience, 2001.


\bibitem{ree93}
B.A.~Reed.  Problem session on parity problems (Public communication).
{\it DIMACS Workshop on Perfect Graphs}, Princeton University, New
Jersey, 1993.



\end{thebibliography}
\end{document}